\documentclass[lettersize,journal]{IEEEtran}
\usepackage[utf8]{inputenc} 
\usepackage[T1]{fontenc}    
\usepackage{url}            
\usepackage{booktabs}       
\usepackage{amsfonts}       
\usepackage{amsmath}
\usepackage{amssymb}
\usepackage{mathtools}
\usepackage{nicefrac}       
\usepackage{microtype}      
\usepackage{xcolor}         
\usepackage{subfiles}

\usepackage{rotating}
\usepackage{multirow}
\usepackage{makecell}
\usepackage[normalem]{ulem}
\useunder{\uline}{\ul}{}

\newcommand{\midsepremove}{\aboverulesep = 0.3025mm \belowrulesep = 0.492mm}
\newcommand{\midsepdefault}{\aboverulesep = 0.605mm \belowrulesep = 0.984mm}

\newcommand{\norm}[1]{\left\|#1\right\|}
\newcommand{\x}{\pmb{x}}
\newcommand{\scm}[1]{\pmb{s}(#1; \theta)}
\newcommand{\vm}[1]{\pmb{v}(#1; \phi)}

\newcommand{\scmf}[2]{\mathcal{E}^{#1}_S(#2; \theta_1)}
\newcommand{\vmf}[2]{\mathcal{E}^{#1}_L(#2; \phi_1)}

\newcommand{\dt}[1]{\textcolor{black}{#1}}

\usepackage{amsmath,amsfonts}
\usepackage{algorithmic}
\usepackage{algorithm}
\usepackage{array}
\usepackage[caption=false,font=normalsize,labelfont=sf,textfont=sf]{subfig}
\usepackage{textcomp}
\usepackage{stfloats}
\usepackage{url}
\usepackage{verbatim}
\usepackage{graphicx}
\usepackage{cite}
\begin{document}

\title{Hybrid Long and Short Range Flows for Point Cloud Filtering}

\author{Dasith~de~Silva~Edirimuni,~
        Xuequan~Lu,~\IEEEmembership{Senior~Member,~IEEE,}
        Ajmal~Saeed~Mian,~\IEEEmembership{Senior~Member,~IEEE,}
        Lei~Wei,~
        Gang~Li,~\IEEEmembership{Senior~Member,~IEEE,}
        Scott~Schaefer,~
        and~Ying~He
\IEEEcompsocitemizethanks{\IEEEcompsocthanksitem D. de Silva Edirimuni, X. Lu and A. Mian are with the Department of Computer Science and Software Engineering, University of Western Australia, Crawley, Western Australia, 6009, Australia (e-mail: \{dasith.desilva, bruce.lu, ajmal.mian\}@uwa.edu.au). L. Wei and G. Li are with the School of Information Technology, Deakin University, Waurn Ponds, Victoria, 3216, Australia (e-mail: \{lei.wei, gang.li\}@deakin.edu.au). S. Schaefer is with the Department of Computer Science and Engineering, Texas A\&M University, College Station, TX, USA (e-mail: schaefer@cse.tamu.edu). Y. He is with the College of Computing and Data Science, Nanyang Technological University, Singapore (e-mail: yhe@ntu.edu.sg).}
\thanks{Manuscript received Month Day, Year; revised Month Day, Year.}
\thanks{(Corresponding author: Xuequan Lu.)}}

\markboth{Journal of \LaTeX\ Class Files,~Vol.~14, No.~8, August~2021}%
{Shell \MakeLowercase{\textit{et al.}}: A Sample Article Using IEEEtran.cls for IEEE Journals}

\IEEEpubid{0000--0000/00\$00.00~\copyright~2021 IEEE}

\maketitle

\begin{abstract}
Point cloud capture processes are error-prone and introduce noisy artifacts that necessitate filtering/denoising. Recent filtering methods often suffer from point clustering or noise retaining issues. In this paper, we propose Hybrid Point Cloud Filtering (\textbf{HybridPF}) that considers both short-range and long-range filtering trajectories when removing noise. 
It is well established that short range scores, given by $\nabla_{x}\log p(x_t)$, may provide the necessary displacements to move noisy points to the underlying clean surface. By contrast, long range velocity flows approximate constant displacements directed from a high noise variant patch $x_0$ towards the corresponding clean surface $x_1$. Here, noisy patches $x_t$ are viewed as intermediate states between the high noise variant and the clean patches. Our intuition is that long range information from velocity flow models can guide the short range scores to align more closely with the clean points. In turn, score models generally provide a quicker convergence to the clean surface. Specifically, we devise two parallel modules, the ShortModule and LongModule, each consisting of an Encoder-Decoder pair to respectively account for short-range scores and long-range flows. We find that short-range scores, guided by long-range features, yield filtered point clouds with good point distributions and convergence near the clean surface. We design a joint loss function to simultaneously train the ShortModule and LongModule, in an end-to-end manner. Finally, we identify a key weakness in current displacement based methods, limitations on the decoder architecture, and propose a dynamic graph convolutional decoder to improve the inference process. Comprehensive experiments demonstrate that our HybridPF achieves state-of-the-art results while enabling faster inference speed. 
\end{abstract}


\section{Introduction}
\IEEEPARstart{P}oint clouds are sets of 3D coordinate data obtained through numerous sensors such as RGB-D cameras and LiDAR scanners. They are widely used in 3D vision, modelling and graphics tasks \cite{Kim-3D-Printing, Urech-Urban-Planning, Luo-Pillar-Motion}, due to their ease of use and compact nature. However, sensor limitations and environmental factors can lead to noisy artifacts in the captured point clouds. Therefore, filtering/denoising is a fundamental point cloud pre-processing task that facilitates subsequent tasks, such as mesh reconstruction \cite{RIMLS-Oztireli, Fleishman-Bilateral}, registration \cite{Fung-SemReg}, 3D modelling \cite{Kim-3D-Printing} and scene understanding \cite{Liao-Kitti360}. 

\IEEEpubidadjcol

Traditional filtering methods generally rely on handcrafted features to estimate displacements which move the points back towards the clean surface. Most traditional methods require normal information \cite{Digne-Bilateral,Mattei-MRPCA,Hu-GLR}, necessitating an additional normal estimation step. However, at high noise levels, such methods become ineffective as the normal estimation and filtering steps both exhibit high sensitivity to noise. The introduction of powerful neural network architectures \cite{Qi-PointNet,Qi-PointNet++,Wang-DGCNN} for encoding latent representations of 3D point clouds makes it possible to develop encoder-decoder models that generate better filtered results \cite{Rakotosaona-PCN, Luo-Score-Based-Denoising, Edirimuni-IterativePFN} compared to traditional methods which use handcrafted features.

\begin{figure}[!tp]
    \centering
    \includegraphics[width=\linewidth]{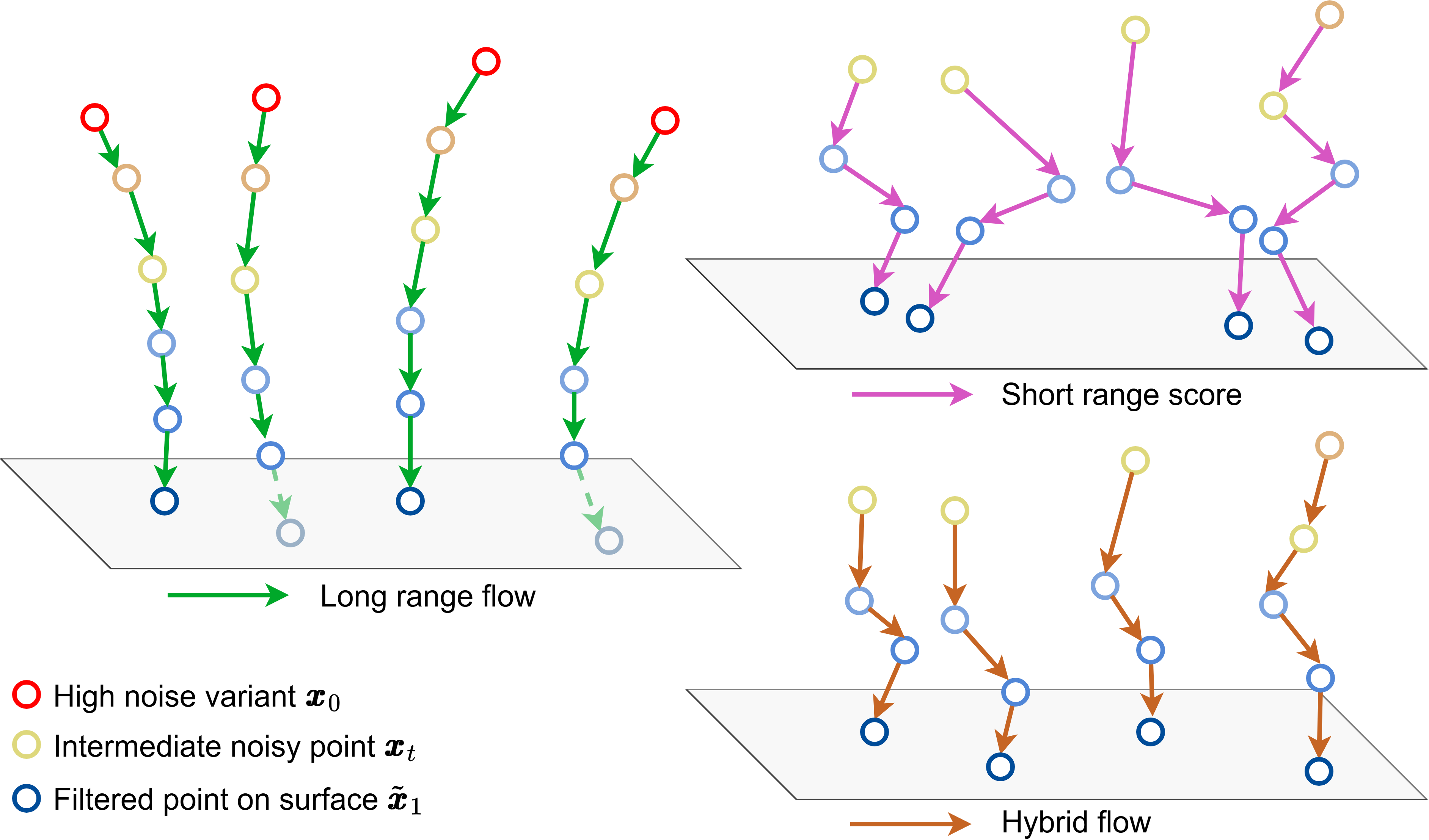}
    \caption{We visualize the hybrid flows which recover better distributions of points. Long range flow methods tend to overshoot/undershoot the true surface. Meanwhile, current score methods suffer from clustering and poorly distributed filtered points. Our hybrid method recovers the clean surface, while retaining a good distribution of points.}
    \label{fig:filtering-trajectories}
    \vspace{-4mm}
\end{figure}

In recent years, score matching based filtering methods \cite{Luo-Score-Based-Denoising, Chen-DeepPSR} have risen in popularity due to their strong filtering ability. The displacement based method IterativePFN \cite{Edirimuni-IterativePFN} draws parallels between displacement and score-matching methods to internally model the iterative filtering process. Such score and displacement based methods are inspired by a stochastic differential equation \cite{Song-Score-Matching}, that describes the Brownian diffusion process:
\begin{equation}
  \label{eq:forward-langevin}
  \text{d}\x_t = \pmb{\mu}(\x_t)\text{d}t + \sigma(t)\text{d}\pmb{w},
\end{equation}
where $\pmb{\mu}(\x_t), \sigma(t)$ are the drift and diffusion coefficients and $\pmb{w}$ is the standard Weiner process. In particular, $\pmb{w}$ introduces stochasticity \cite{Song-Score-Matching, Song-SDE-PF-ODE} to the trajectory of particles. Eq.~\eqref{eq:forward-langevin} governs the manner in which points diffuse from an initial clean state to a noisy variant. The corresponding reverse Langevin equation that defines the filtering process is:
\begin{equation}
  \label{eq:backward-langevin}
  \text{d}\x_t = [\pmb{\mu}(\x_t) - \sigma(t)^2\underbrace{\nabla_{\x} \log p(\x_t)}_{\scm{\x_t}}]\text{d}t + \sigma(t)\text{d}\bar{\pmb{w}},
\end{equation}
where $\bar{\pmb{w}}$ is a Weiner process and the score model $\scm{\x_t}$, with parameters $\theta$, approximates $\nabla_{\x} \log p(\x_t)$. 
While early score matching methods brought substantial improvements over previous state-of-the-art methods, they needed a high number of iterations to recover the underlying clean surface \cite{Luo-Score-Based-Denoising}. This was addressed by more recent work \cite{Edirimuni-IterativePFN}. Moreover, the filtered points obtained by score matching methods \cite{Luo-Score-Based-Denoising, Edirimuni-IterativePFN, Chen-DeepPSR} tend to cluster together, indicating the difficulty in recovering the true point distribution (see Fig.~\ref{fig:filtering-trajectories}).

\begin{figure}[!tp]
    \centering
    \includegraphics[width=\linewidth]{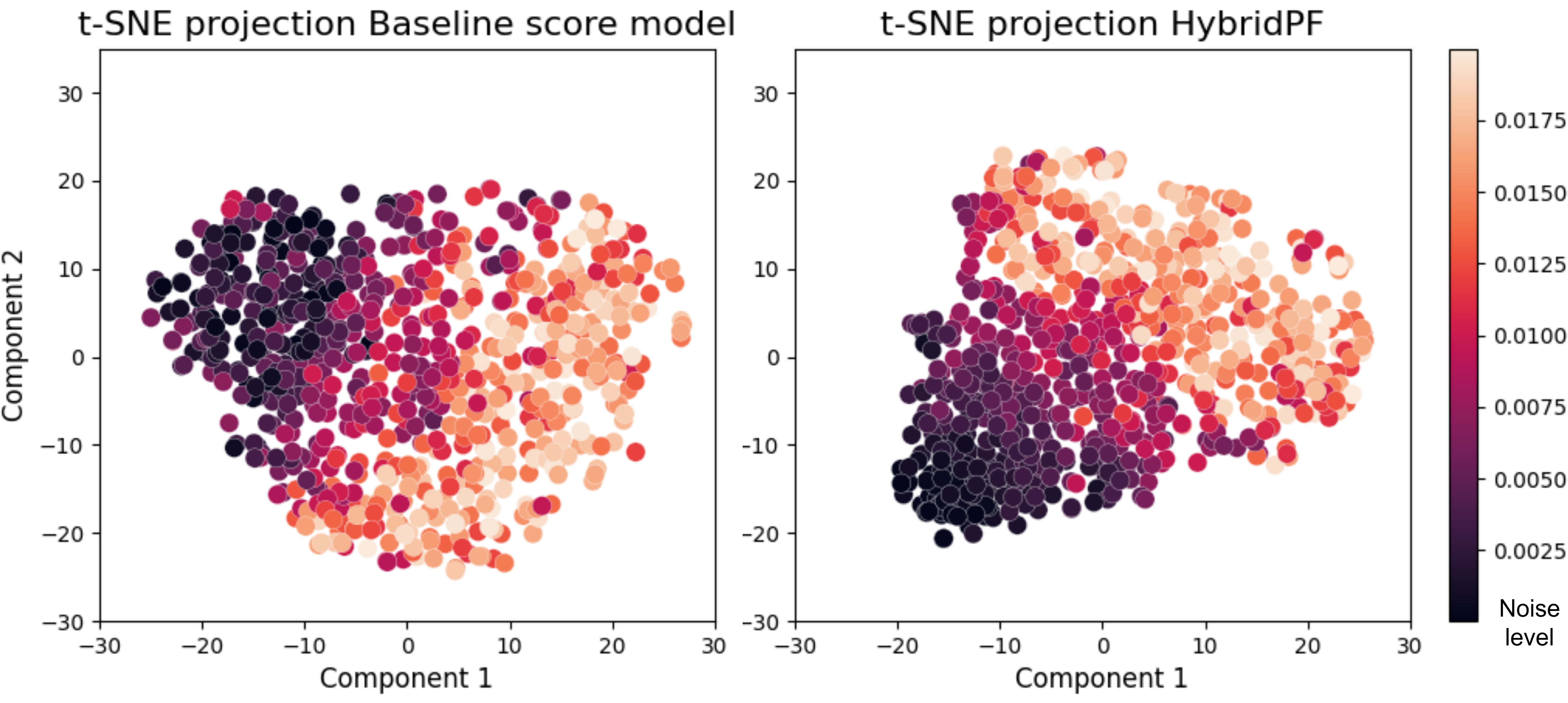}
    \caption{We use t-SNE to visualize features of a {\em single} point, at different noise levels. At lower noise levels, we expect features to be similar (i.e., small distance between darker points). As noise increases, features are less consistent. However, the encoder should ideally distinguish features across noise levels (clearer separation of colors). The baseline score model is sub-optimal as features at lower noise (dark) are mixed with features at medium noise (light). 
    By contrast, HybridPF produces more consistent low noise features, 
    i.e. the darker/lower the points/noise, the closer they are.}
    \label{fig:tsne}
\end{figure}

More recently, Rectified Flow (Reflow) based methods \cite{Liu-RectifiedFlow, Wu-PointStraightFlow} introduced constant long range flows for 2D and 3D generation which satisfy the ordinary differential equation $\text{d}\x_t = \pmb{v}(\x_t)\text{d}t$. This deterministic approach has been adapted for point cloud filtering \cite{Edirimuni-StraightPCF}. Here, a noisy point $\x_t$ is modelled as an intermediate state, interpolated between a noisy data $\x_0~\sim\mathcal{N}(0,\sigma_H)$ and the clean point $\x_1$, where $\sigma_H$ is the standard deviation for the highest noise scale. Subsequently, the constant flow $\pmb{v}(\x_t)$ is approximated by a neural network $\vm{\x_t}$, with parameters $\phi$. More precisely, the network infers the constant vector $ \x_1-\x_0$. Thereafter, using an ODE solver, noisy points are displaced along constant, straight paths towards the underlying surface. However, the flow is approximate and straightness is not guaranteed. Generally, this method requires sufficiently small discretization steps and trajectories may be curved, as seen in Fig.~\ref{fig:filtering-trajectories}. While StraightPCF obtains better point distributions, due to the long range information $\x_1-\x_0$, filtered points may still overshoot or undershoot the clean surface. Furthermore, StraightPCF requires a lengthy 3-stage training process to train and refine the coupled VelocityModule mechanism. At test time, higher noise levels require higher iteration numbers, which increases latency.

Finally,  A secondary concern among both score and reflow based methods is the limitation of the decoder when decoding high dimensional features into displacements in 3D Euclidean space. Current methods such at IterativePFN and StraightPCF \cite{Edirimuni-IterativePFN, Edirimuni-StraightPCF} use a simple stack of fully connected layers for decoding the point features. This decoder architecture is sub-optimal. We surmise that a more effective decoder architecture may be developed using dynamic graph convolutional layers \cite{Wang-DGCNN}. The dynamic graph convolution pays attention to each point's local neighborhood, in latent space, when decoding the final score displacement. This directly contrasts with current methods where fully connected layers make no attempt to discriminate between point neighbors during the decoding process. Based on the above observations of current limitations, we derive the following insights: 1) Our key insight is that the long range flow information can guide the short range stochastic flow, i.e., scores. 2) Improving decoder architecture, using graph convolutional layers, can significantly improve the filtering process.

In this paper, we propose a unique approach for point cloud filtering, that considers both short-range and long-range information, namely Hybrid Point Cloud Filtering (HybridPF). To this end, we design a network that consists of two parallel modules, ShortModule and LongModule, respectively indicating short-range and long-range flows. Each module is composed of a Encoder-Decoder structure. The LongModule is trained to infer long-range, constant flows $\x_1-\x_0$. 
The ShortModule subsequently infers short range scores conditioned on long range features produced by the LongModule's encoder. This provides the ShortModule with long range information about the relationship between $\x_t$, $\x_0$ and $\x_1$, which improves the estimated score's accuracy. The impact of this long range conditioning on the ShortModule is visualized in Fig.~\ref{fig:tsne}.
In summary, our contributions are: 
\begin{itemize}
  \item We propose HybridPF, that takes both short range and long range information into account for filtering points along stochastic trajectories quickly, and effectively. 
  \begin{itemize}
      \item Based on the intuition that long range information can guide the short range flow, we design a joint loss function to train the ShortModule and LongModule, simultaneously, in an end-to-end manner. 
  \end{itemize}
  \item Current decoder architectures are limited, for displacement based methods. We propose using dynamic graph convolutional layers to improve the decoder, which leads to better filtering results. 
\end{itemize}
We conduct comprehensive experiments on Kinect data, LiDAR data and synthetic data, and demonstrate that our method yields state-of-the-art results, while improving runtime efficiency over recent methods such as PDFlow \cite{Mao-PDFlow} and IterativePFN \cite{Edirimuni-IterativePFN}. 


\section{Related Work}
\label{sec:review}
\paragraph{Conventional filtering methods:} Early filtering methods built on the work of Levin \cite{MLS-Levin} who proposed the Moving Least Squares method to obtain surface information such as orientation by estimating point normals. These include the works of Alexa \emph{et al.} \cite{Alexa--MLS-PSS}, Adamson and Alexa \cite{IMLS-Adamson} and Guennebaud and Gross \cite{APSS-Guennebaud}. Alternatives to MLS use PCA analysis \cite{Hoppe-PCA} or n-jet fitting \cite{Cazals-Jet} to generate  point normals. Point normals provide surface orientation and curvature information that can be exploited within handcrafted features to filter noisy points, as explored by Digne \cite{Digne-Similarity}. Moreover, Digne and de Franchis \cite{Digne-Bilateral} proposed a bilateral filtering \cite{Fleishman-Bilateral} based point set filtering mechanism that moves points along normal directions. Other conventional methods include the $L_1$ and $L_0$ minimization schemes of Avron \emph{et al.} \cite{Avron-L1} and Sun, Schaefer and Wang \cite{Sun-L0} to obtain point normals and use them within position update mechanisms. Mattei and Castrodad introduced the Moving Robust Principal Component Analysis (MRPCA) \cite{Mattei-MRPCA} while Lu \emph{et al.} developed the Low Rank Matrix Approximation \cite{Lu-Low-Rank} method to estimate normals which are later used for filtering. Hu \emph{et al.} proposed the Graph Laplacian Regularization (GLR) \cite{Hu-GLR} method to filter points based on the kNN graph for the point set.  

Contrary to normal based methods, Lipman \emph{et al.} introduced the Locally Optimal Projection (LOP) \cite{Lipman-LOP} to resample point clouds. First they downsample and regularize the point cloud; however, this requires an additional upsampling step to recover the same point resolution. The works of Huang \emph{et al.} (WLOP) \cite{Huang-WLOP} and Preiner \emph{et al.} (CLOP) \cite{Preiner-CLOP} improved  the vanilla LOP implementation. Nevertheless, both PCA and resampling based methods showed significant sensitivity to noise due to errors propagated through the normal estimation and downsampling steps, respectively. Furthermore, resampling also resulted in the smoothening of geometric features. 

\begin{figure*}[!tp]
    \centering
    \includegraphics[width=\linewidth,trim=4 4 4 4,clip]{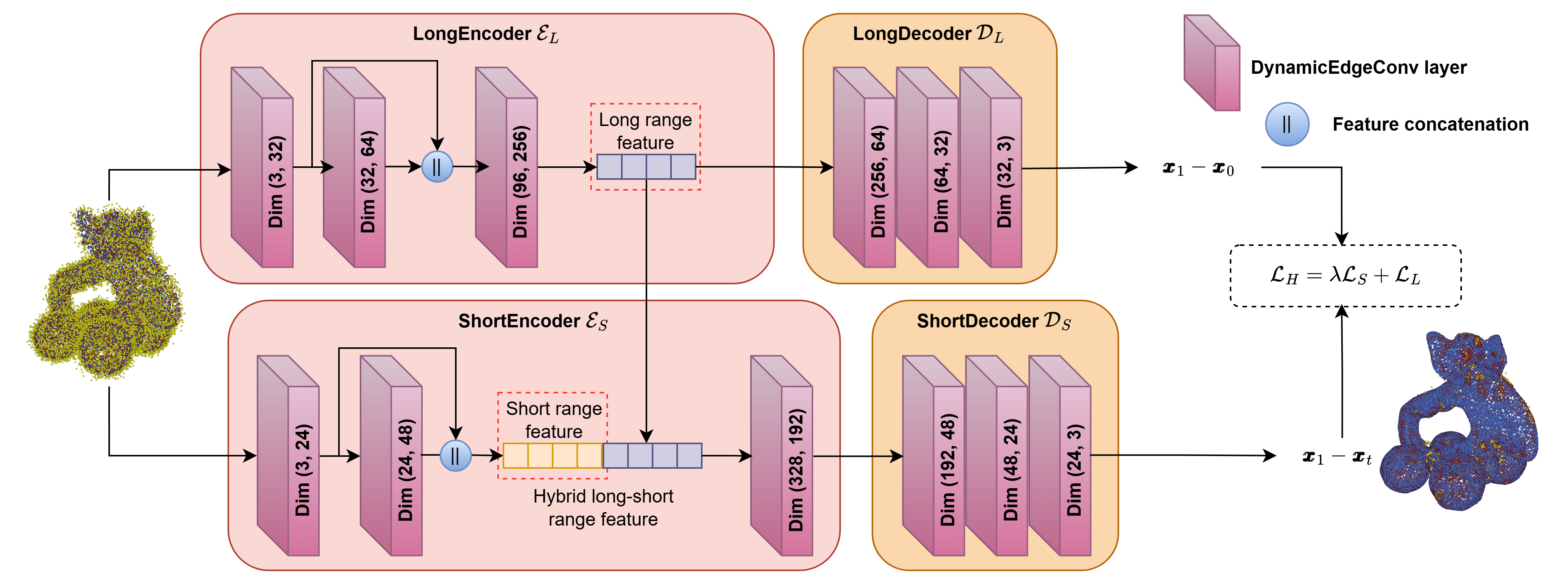}
    \caption{Our HybridPF architecture. It comprises of a LongModule that infers long range flow and a ShortModule that infers short range flow. During training, we optimize the joint loss $\mathcal{L}_H$ using the output of both the LongModule and ShortModule. However, during inference, we discard the LongDecoder $\mathcal{D}_L$ and use the latent representations from the LongEncoder $\mathcal{E}_L$ to guide the short range flow estimation of the ShortModule. }
    \label{fig:network}
\end{figure*}

\vspace{1mm}
\noindent {\bf Deep learning based filtering methods:} 
Given the pronounced drawbacks of conventional methods in both normal estimation and point filtering, more recent works have focused on deep learning to generate richer latent representations which can be more effectively decoded to provide filtered displacements. Early works rely on convolutional neural networks for regressing normals and filtered displacements from 2D height maps \cite{Lu-Deep-Feature-Preserving, Roveri-PointProNets}. In this regard, PointNet \cite{Qi-PointNet} and PointNet++ \cite{Qi-PointNet++} pioneered the way for directly generating robust 3D features directly from point sets. PointNet was first used as a backbone in the encoder-decoder model PointCleanNet (PCN) \cite{Rakotosaona-PCN} which decodes the rich latent features from the PointNet encoder using a simple MLP based decoder. The decoder output is a filtered displacement that shifts the noisy point from the original noisy patch towards the clean surface. This paradigm of point filtering was immediately successful and PointNet received widespread adoption as a backbone, with Pointfilter \cite{Zhang-Pointfilter} of Zhang \emph{et al.} and CFilter \cite{Edirimuni-Contrastive-Joint-Learning} of Edirimuni \emph{et al.} further refining on the displacement based scheme. CFilter \cite{Edirimuni-Contrastive-Joint-Learning}, in particular, jointly infers point positions and point normals. They use the estimated normals in a post step to refine the initial filtered positions. Pointfilter \cite{Zhang-Pointfilter} is a simple yet powerful framework, that embeds clean normal information into training only. A significant shortcoming of these methods, however, is the large size of their networks. Furthermore, these methods use large patches of approximately 500 points to filter a single, central point which leads to high computation times and inefficient use of resources.

While previous direct point set convolution architectures were computationally less efficient, DGCNN \cite{Wang-DGCNN} of Wang \emph{et al.} instead proposed a graph convolution network (GCN) based approach where a k-nearest neighbor graph was constructed from the point set. Each point is a graph node, with the initial point position as the node feature. This GCN approach permits simultaneously convolving all nodes to generate rich node features that can then be decoded in regression tasks. Among the first to adopt a GCN approach to point cloud filtering were Pistilli \emph{et al.} who proposed GPDNet \cite{Pistilli-GPDNet}. Simultaneously, Luo and Hu introduced the resampling based DMRDenoise \cite{Luo-DMRDenoise} which suffered from similar issues to traditional resampling techniques. However, a significant step towards improving filtering was made by Luo and Hu in their work ScoreDenoise \cite{Luo-Score-Based-Denoising} which exploited the gradient log of the probability distribution of point positions as estimates of filtered displacements. This was further extended by Chen \emph{et al.} in their DeepPSR approach that combines score matching with graph Laplacian regularization. These gradient based approaches were pioneered by Song \emph{et al.} \cite{Song-Score-Matching,Song-SDE-PF-ODE} and have seen applications in point cloud generation \cite{Cai-ShapeGF, Luo-Diffusive}. Meanwhile, de Silva Edirimuni \emph{et al.} drew parallels between displacement and score matching based mechanisms and proposed IterativePFN \cite{Edirimuni-IterativePFN}, that models iterative filtering internally, using their IterationModule. This alleviates the need to iteratively apply filtering at test time but results in a large network with over $3M$ parameters.

In contrast to IterativePFN, Chen \emph{et al.} proposed RePCDNet \cite{Chen-RePCD} that exploits a recurrent neural network architecture to model iterative filtering. However, the network must be iteratively applied during test time. Mao \emph{et al.} proposed PDFlow \cite{Mao-PDFlow}, which recovers clean representations of input noisy points via normalizing flows. This was further extended in their subsequent work, PD-LTS \cite{Mao-PD-LTS}, to combine graph convolution together with invertible neural networks. While score matching infers the gradient log of the probability distribution, a short range flow from $\x_t$ to the clean point $\x_1$, Reflow \cite{Liu-RectifiedFlow} models $\x_t$ as intermediate states in an interpolation between a high noise variant $\x_0$ and the clean point $\x_1$. This was first applied to point cloud generation and was adapted to point cloud filtering by Edirimuni \emph{et al.} \cite{Edirimuni-StraightPCF}. 

\section{Method}
\label{sec:method}
Given a noisy point cloud $P_{\pmb{X}} = \{x^i~|~x^i \in \mathbb{R}^3\}$, the goal of point cloud filtering (also known as denoising) is to recover the original clean point cloud $P_{\pmb{Y}} = \{y^i~|~y^i \in \mathbb{R}^3\}$. $P_{\pmb{X}}$ is a noisy variant of $P_{\pmb{Y}}$ with $P_{\pmb{X}} = P_{\pmb{Y}} + \sigma\xi$ where $\xi\sim\mathcal{N}(0,1)$ and $\sigma\in\mathbb{R}$. In practice, since filtering a large point cloud is inefficient, it is standard practice to filter point cloud patches $\pmb{X}=\{\pmb{x}^i|\pmb{x}^i\in kNN(\pmb{x}^r, \mathcal{P}_{\pmb{X}}, k)\}$ centered around $\pmb{x}^r$, which are reference points sampled from $P_{\pmb{X}}$. The ultimate aim is to recover the corresponding clean points $\pmb{Y}=\{\pmb{y}^i|\pmb{y}^i\in \mathcal{P}_{\pmb{Y}}\}$. Once filtered, the patches are recombined to form the complete filtered point cloud. Methods such as IterativePFN \cite{Edirimuni-IterativePFN} and ScoreDenoise \cite{Luo-Score-Based-Denoising} displace noisy points along stochastic trajectories towards the clean surface and filter all points within a patch simultaneously. 

\subsection{Decoder architecture}
DGCNN based graph convolution has been widely adopted for encoding point features. 
IterativePFN is a prominent example, which has inspired other methods. The encoder consumes a graph $\mathcal{G}=(V, E)$ where $V$ and $E$ are the vertex set and edge set, respectively. Each point in the patch is a vertex in the graph and an edge is formed between each point and its 32 nearest neighbors. During message-passing, the $i$-th vertex's features , $\mathbf{h}^{l+1}_i$, at the $l+1$-th layer is obtained by $ \mathbf{h}^{l+1}_i = \text{MLP}(\mathbf{h}^{l}_i) + \Box_{j:(i,j) \in E} \text{MLP}(\mathbf{h}^{l}_i \, \Vert \, \mathbf{h}^{l}_j - \mathbf{h}^{l}_i)$, where $\Box_{j:(i,j) \in E}$ is the element-wise max pooling operator and $(\cdot\,\Vert\,\cdot)$ represents concatenation. 

However, a key research gap is the decoder. IterativePFN and other variants use a sequence of fully connected layers to decode the feature. Unlike the encoder, that aggregates neighborhood information according to the dynamic graph, current decoders do not consider any topological information in this high dimensional feature space. Therefore, we introduce a decoder architecture comprising Dynamic EdgeConv layers, instead of simple fully connected layers (see Fig.~\ref{fig:network}). During the decoding process, the decoder dynamically builds a graph of point features and their neighbors in latent space. During message-passing, this topological information is considered when aggregating graph signals. The impact of the graph based decoder is discussed in Sec.~\ref{sec:ablation}.

\subsection{Stochastic short range flows}
The noisy point cloud patch $\pmb{X}$ can be viewed as an intermediate state $\pmb{X}_t$ in a filtering process that moves it towards the clean surface $\pmb{Y} = \pmb{X}_1$ across time steps $t \in [0,1]$. The filtering process of score matching 
is governed by Eq.~\eqref{eq:backward-langevin} where the quantity $\nabla_{\x} \log p(\x_t)=(\x_1-\x_t)/\sigma^2$ shifts noisy points towards the clean surface. This quantity is approximated by the score network $\scm{\cdot}$. We identify this as a short range flow trajectory from $\x_t$ to $\x_1$. Similar to previous works \cite{Luo-Score-Based-Denoising,Edirimuni-IterativePFN}, we model a vanilla ShortModule as an encoder-decoder network $\scm{\x_t}=\mathcal{D}_S(\scmf{}{\x_t}; \theta_2)$ where the ShortModule parameters $\theta=\{\theta_1, \theta_2\}$. This ShortModule only attempts to infer short range flow trajectories from $\x_t$ to $\x_1$ and is typically trained using the following loss objective:
\begin{equation}
  \label{eq:scm-loss}
  \mathcal{L}^{L_2}_S = \mathbb{E}_{t\sim\mathcal{U}(0,1)}\left[\norm{\scm{\x_t} - \left(\x_1 - \x_t\right)}^2_2\right].
\end{equation}
Therefore, $\scm{\x_t}$ infers the displacement required to move the noisy point to the clean surface. That is, $\Tilde{\x}_1 = \x_t + \scm{\x_t}$. Given the filtered point $\Tilde{\x}_1$, based on the ShortModule output, we may also use the Earth Mover Distance (EMD) loss objective, similarly to other methods such as PDFlow and PD-LTS \cite{Mao-PDFlow}, to train the ShortModule.
\begin{equation}
  \label{eq:scm-emd-loss}
  \mathcal{L}^{EMD}_S = \min_{\Omega:\Tilde{\pmb{X}}_1\rightarrow\pmb{X}_1}\sum_{\Tilde{\x}_1\in\Tilde{\pmb{X}}_1}\norm{\tilde{\x}_1-\Omega(\tilde{\x}_1)}_2,
\end{equation}
where $\Omega$ is a bijection from the filtered patch to the clean patch. Short range score matching methods only capture the information $\x_t$ and $\x_1$ and previous works suffer from point clustering. However, combining long range information between $\x_0$, $\x_t$ and $\x_1$ may lead to better filtered point distributions and motivates our Hybrid filtering approach.

\subsection{Constant long range flows}
Unlike score matching methods, Reflow methods \cite{Liu-RectifiedFlow,Wu-PointStraightFlow,Edirimuni-StraightPCF} simulate the deterministic ODE process:
\begin{equation}
  \label{eq:velocity-flow-equation}
  \text{d}\x_t = \pmb{v}(\x_t)\text{d}t,
\end{equation}
where the rate of change of position w.r.t. time is simply a \emph{velocity}. Moreover, they model noisy points as intermediate states $\x_t$ at times $t\in[0,1]$, such that,
\begin{equation}
    \label{eq:vm-linear-interpolation}
    \pmb{x}_t = (1-t)\pmb{x}_0 + t\pmb{x}_1.
\end{equation} 
Thereafter, to ensure a constant flow velocity, they simply optimize:
\begin{equation}
  \label{eq:ode-solution}
  \min_{v}\int^1_0\mathbb{E}_{t\sim\mathcal{U}(0,1)}\left[\norm{\pmb{v}(\x_t) - (\x_1 - \x_0)}^2_2\right]\text{d}t,
\end{equation}
as the solution $\x_1-\x_0$, based on the linear interpolation assumption, satisfies this \emph{velocity-flow} equation (Eq.~\eqref{eq:velocity-flow-equation}). In practice, they approximate $\pmb{v}(\x_t)$ using an encoder-decoder model $\vm{\x_t}$ that infers the constant displacement $\x_1-\x_0$. We identify $\x_1-\x_0$ as a long range flow which contains information relating to the clean point $\x_1$ and the high noise variant $\x_0$. We observe this knowledge of $\x_0$ relative to $\x_1$ encourages the filtered points to be better distributed, with fewer clustering artifacts.

Inspired by these Reflow methods, our LongModule models noisy point cloud patches $\pmb{X}$ as intermediate states in a linear interpolation between a high noise variant patch $\pmb{X}_0\sim\Pi_0$ and the clean patch $\pmb{X}_1\sim\Pi_1$. High noise variants are constructed by adding noise at a high noise scale $\sigma_H=2\%$ of the point cloud's bounding sphere radius such that $\pmb{X}_0 = \pmb{X}_1 + \sigma_H\xi \land \xi\sim\mathcal{N}(0,1)$. 
Points $\x^i_0\in\pmb{X}_0$ and $\x^i_1\in\pmb{X}_1$ are independent and identically distributed samples following the $\Pi_0$ and $\Pi_1$ distributions, respectively. Now, to approximate the \emph{velocity-flow} equation (Eq.~\eqref{eq:velocity-flow-equation}) we use the LongModule, an encoder-decoder model $\vm{\x_t}=\mathcal{D}_L(\mathcal{E}_L(\x_t; \phi_1); \phi_2)$ where the model parameters $\phi=\{\phi_1, \phi_2\}$. The LongModule is trained using the following training objective:
\begin{equation}
  \label{eq:vm-loss}
  \mathcal{L}_L = \mathbb{E}_{t\sim\mathcal{U}(0,1)}\left[\norm{\vm{\x_t} - \left(\x_1 - \x_0\right)}^2_2\right].
\end{equation}

\subsection{HybridPF: short and long range flows for point filtering}
The LongModule, in isolation, only infers the long range flow $\x_1-\x_0$. As such, points may overshoot or undershoot the clean surface, and may not converge. Our intuition is that the long range information from a LongModule can help inform the short-range flow from the ShortModule, i.e., scores, which in turn may lead to better filtering results while converging to clean surface. The guidance from the LongModule guides the overal trajectory and yields better filtered point distributions. The ShortModule and the LongModule combined, illustrated in Fig.~\ref{fig:network}, form our HybridPF. In particular,
\begin{enumerate}
  \item a LongModule that infers a long range flow feature $\vmf{}{\x_t}$ corresponding to the long range displacement $\vm{\x_t} = \x_1-\x_0$. The LongModule decoder is only used during training and is discarded once training has been completed, 
  \item a ShortModule that infers a short range flow, i.e., scores, $\scm{\x_t, \vmf{}{\x_t}} = \x_1 - \x_t$ where the LongRangeEncoder's feature output from the final layer is used as a conditional prior to guide the score estimation.
\end{enumerate}
The key feature of this ShortModule is the use of the point features generated by the LongModule encoder, $\vmf{}{\x_t}$. Now the training objective for our full HybridPF network is given by:
\begin{equation}
  \label{eq:hybrid-loss}
  \mathcal{L}_H = \lambda \mathcal{L}^{EMD}_S + \mathcal{L}_L,
\end{equation}
where $\lambda$ is a hyper-parameter that weights the contribution of the short range loss term.
\subsection{Filtering objective} 
During test time, we apply the filtering process iteratively. Consequently, the ShortModule filters noisy points using the following filtering objective:
\begin{equation}
  \label{eq:score-filtering-objective}
  \Tilde{\x}_{(\hat{t}+1)/N} = \Tilde{\x}_{\hat{t}/N} + \alpha\scm{\Tilde{\x}_{\hat{t}/N}, \vmf{}{\Tilde{\x}_{\hat{t}/N}}},
\end{equation}
with $\alpha=0.8$, the discretization step, chosen empirically. The filtering objective is applied $N=4$ times along integer time steps $\hat{t}$. We provide more details on tuning $\alpha$ and $N$ in the supplementary. The full position update is given by:
\begin{equation}
  \label{eq:score-full-filtering-objective}
  \Tilde{\x}_{1} = \Tilde{\x}_0 + \alpha\sum^N_{\hat{t}=0}\scm{\Tilde{\x}_{\hat{t}/N}, \vmf{}{\Tilde{\x}_{\hat{t}/N}}}.
\end{equation}
The starting time for the filtering process is not known. Therefore, during inference, we set $\Tilde{\x}_0=\x_t$ where $\x_t$ is the input noisy patch.

\section{Experiments}
\label{sec:experiments}
\subsection{Implementation and datasets}
\textbf{Implementation.} We train and test the proposed HybridPF architecture on a NVIDIA Geforce RTX 3090 GPU with PyTorch 2.0.1 and CUDA 12.2. We train our method for 600K training steps using the Adam optimizer and learning rate of $1\times 10^{-4}$. For fair comparisons, all compared methods are trained only on PUNet data with Gaussian noise. We quote results from each respective method, where possible.

\textbf{Metrics.} For the quantitative evaluation, we use two standard metrics: 1) the Chamfer distance (CD) which measures the similarity between the filtered point cloud and the clean point cloud and 2) the Point2Mesh (P2M) distance which measures the similarity between the filtered point cloud and the ground truth mesh. In particular, we use PyTorch3D \cite{Ravi-PyTorch3D} implementations of these metrics, consistent with \cite{Luo-Score-Based-Denoising, Edirimuni-IterativePFN}. P2M distance does not effectively penalize point clustering, whereas Chamfer distance ensures that filtered points should be well distributed as the clean point clouds are sparse.

\midsepremove
\begin{table*}[!t]
\setlength{\tabcolsep}{6.8pt}
\centering
\caption{Filtering results on the PUNet and PCNet datasets with Gaussian noise. Our method generalizes well to high noise levels ($\sigma=3\%$) unseen during training. Moreover, on the PCNet dataset, unseen during training, we obtain the best CD and P2M results across most cases. CD and P2M values are multiplied by $10^4$.  }
\begin{tabular}{c|l|cccccc|cccccc}
\toprule
\multicolumn{2}{c|}{Resolution} & \multicolumn{6}{c|}{10K (Sparse)} & \multicolumn{6}{c}{50K (Dense)} \\
\midrule
\multicolumn{2}{c|}{Noise} & \multicolumn{2}{c}{1\%} & \multicolumn{2}{c}{2\%} & \multicolumn{2}{c|}{3\%} & \multicolumn{2}{c}{1\%} & \multicolumn{2}{c}{2\%} & \multicolumn{2}{c}{3\%} \\
\midrule
\multicolumn{2}{c|}{Method} & CD & P2M & CD & P2M & CD & P2M & CD & P2M & CD & P2M & CD & P2M \\
\midrule
\parbox[t]{2mm}{\multirow{6}{*}{\rotatebox[origin=c]{90}{PUNet \cite{Yu-PUNet}}}}
    & Score \cite{Luo-Score-Based-Denoising} & 2.52 & 0.46 & 3.69 & 1.07 & 4.71 & 1.94 & 0.72 & 0.15 & 1.29 & 0.57 & 1.93 & 1.04 \\
    & PDFlow \cite{Mao-PDFlow} & 2.13 & 0.38 & 3.25 & 1.01 & 4.45 & 2.00 & 0.65 & 0.16 & 1.17 & 0.58 & 1.91 & 1.21 \\
    & DeepPSR \cite{Chen-DeepPSR} & 2.42 & 0.33 & 3.44 & 0.82 & 4.17 & 1.35 & 0.65 &  0.08 & 1.026 & 0.33 & 1.39 & \bf 0.58 \\
    & IterativePFN \cite{Edirimuni-IterativePFN} & 2.06 & 0.22 & 3.04 & 0.56 & 4.24 & 1.38 & 0.61 & \bf 0.06 & 0.80 & \bf 0.18 & 1.97 & 1.01  \\
    & StraightPCF \cite{Edirimuni-StraightPCF} & 1.87 & 0.24 & 2.64 & 0.60 & \uline{3.29} & 1.13 &  0.56 & 0.11 & 0.77 & 0.27 & \uline{1.31} & 0.65 \\ 
    & PD-LTS \cite{Mao-PD-LTS} & \bf 1.83 & \bf 0.21 & \uline{2.57} & \bf 0.52 & 3.40 & \bf 1.10 & \uline{0.50} & \bf 0.06 & \uline{0.71} & \uline{0.19} & 1.44 & 0.73 \\ 
\cmidrule{2-14}
    & \bf Ours & \bf 1.83 & \bf 0.21 & \bf 2.55 & \bf 0.52 & \bf 3.23 & \uline{1.11} & \bf 0.49 & \bf 0.06 & \bf 0.70 & 0.20 & \bf 1.27 & \uline{0.61} \\  
\midrule
\parbox[t]{2mm}{\multirow{6}{*}{\rotatebox[origin=c]{90}{PCNet \cite{Rakotosaona-PCN}}}}
    & Score \cite{Luo-Score-Based-Denoising} &  3.37 &  0.83 &  5.13 &  1.20 &  6.78 &  1.94 &  1.07 &  0.18 &  1.66 &  0.35 &  2.49 &  0.66  \\
    & PDFlow \cite{Mao-PDFlow} & 3.24 & 0.61 & 4.55 & 0.97 & 5.93 & 1.44 & 0.97 & 0.15 & 1.65 & 0.42 & 2.45 & 0.57 \\
    & DeepPSR \cite{Chen-DeepPSR} &  3.14 & 1.01 & 4.93 & 1.38 & 6.14 & 1.78 & 0.99 & 0.17 & 1.55 & 0.38 & 2.12 & 0.59 \\
    & IterativePFN \cite{Edirimuni-IterativePFN} & \bf 2.62 & 0.70 & 4.44 & 1.01 & 6.03 & 1.56 & 0.91 & 0.14 & 1.25 & \uline{0.24} & 2.53 & 0.72 \\
    & StraightPCF \cite{Edirimuni-StraightPCF} & \uline{2.75} & \uline{0.54} & \uline{4.05} & \uline{0.79} & \bf 4.92 & \bf 1.09 & 0.88 & 0.14 & \uline{1.17} & 0.26 & \bf 1.82 & \bf 0.45 \\ 
    & PD-LTS \cite{Mao-PD-LTS} & 2.84 & 0.54 & 4.15 & 0.83 & 5.34 & 1.26 & \uline{0.84} & 0.13 & 1.21 & 0.25 & 2.02 & 0.53 \\ 
    \cmidrule{2-14}
    & \bf Ours & 2.79 & \bf 0.47 & \bf 3.97 & \bf 0.70 & \uline{4.99} & \uline{1.12} & \bf 0.79 & \bf 0.10 & \bf 1.12 & \bf 0.20 & \uline{1.91} & \uline{0.48} \\
\bottomrule
\end{tabular}
\label{tab:pu-pc-results}
\end{table*}
\midsepdefault

\begin{figure*}[!t]
    \centering
    \includegraphics[width=\linewidth]{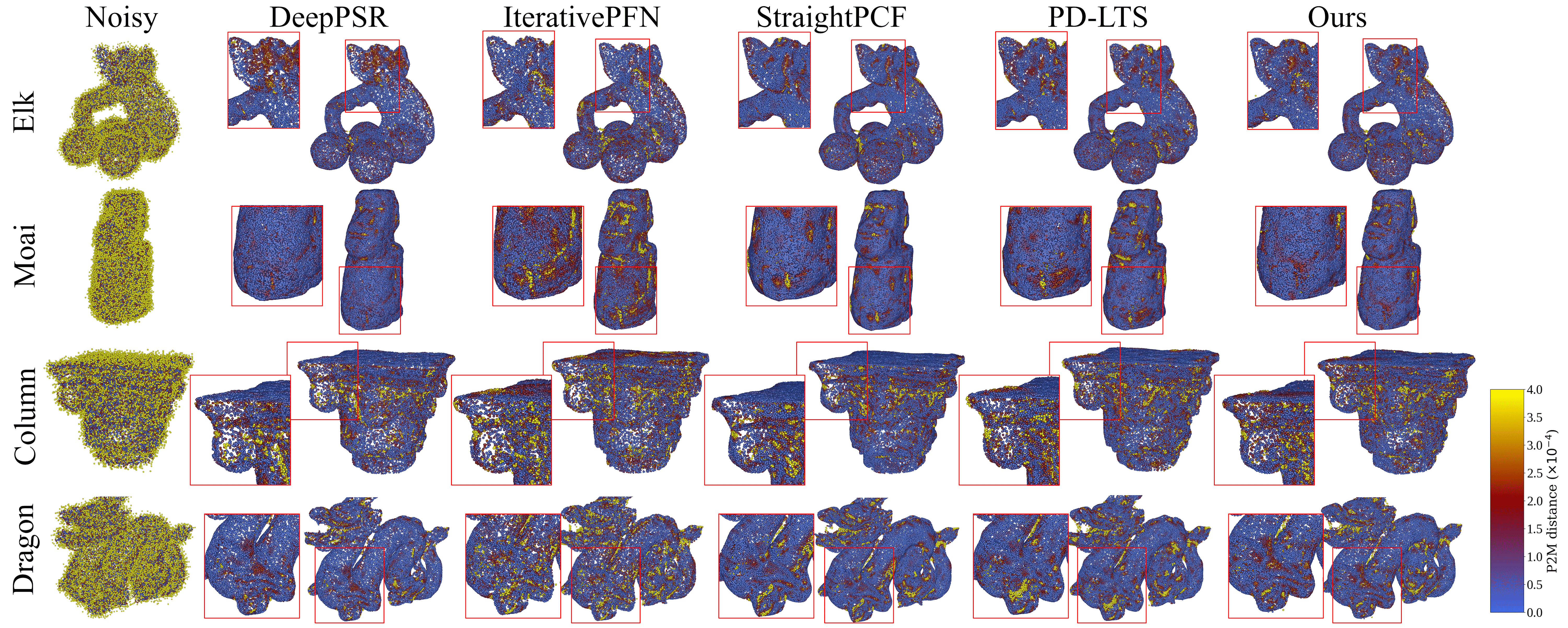}
    \caption{Heatmap visualization of filtered point clouds on PUNet and PCNet data at high resolution and high Gaussian noise (i.e., 50K resolution shapes with $\sigma=3\%$). Blue indicates filtered points are closer to the clean surface. 
    }
    \label{fig:pu-pc-results}
\end{figure*}

\textbf{Synthetic data.} 
Consistent with other recent methods \cite{Edirimuni-IterativePFN, Luo-Score-Based-Denoising}, we train the HybridPF network on the PUNet dataset which consists of 40 synthetic training and 20 synthetic evaluation point clouds. For training, point clouds are sampled at 3 different resolution levels: 10K, 30K and 50K points. Thereafter, Gaussian noise at the highest noise scale $\sigma_H=2\%$ of the bounding sphere radius is added to each point cloud. Intermediate noisy point clouds are then constructed in accordance with the linear interpolation of Eq.~\eqref{eq:vm-linear-interpolation}. At testing time, we evaluate on the PUNet testing set at two different resolutions of 10K and 50K points and noise scales of $1\%$, $2\%$ and $3\%$ of the point cloud's bounding sphere radius. Furthermore, we evaluate results on 10 point clouds of the PCNet dataset \cite{Rakotosaona-PCN}, as provided by \cite{Luo-Score-Based-Denoising}, under the same resolution and noise settings as the synthetic PUNet testing data. 

The \emph{supplementary document} provides additional results on synthetic data for different noise patterns. 

\textbf{Real-world data.}
We also evaluate methods on standard real-world datasets to measure their effectiveness in real-world scenarios. We provide results on 4 scenes of the Paris-Rue-Madame dataset \cite{Serna-Rue-Madame} and 71 point clouds from the Kinect dataset \cite{Wang-Kinect}. As Paris-Rue-Madame comprises real world scans lacking corresponding ground truth data, only qualitative visual results are provided.

\subsection{Evaluation results on synthetic data}
Table~\ref{tab:pu-pc-results} and Fig.~\ref{fig:pu-pc-results} show  quantitative and visual filtering results on the PUNet and PCNet data with synthetic Gaussian noise. 
We observe that while IterativePFN obtains strong P2M results, it performs poorly on the Chamfer distance metric. The converse is true for StraightPCF. Meanwhile, PD-LTS performs well on low noise settings but does sub-optimally on high, unseen noise. Stochastic methods such as IterativePFN and DeepPSR suffer from clustering, as seen in Fig.~\ref{fig:pu-pc-results} where a significant number of holes and large yellowish regions, i.e., points lacking surface convergence, are present on the Elk and Column shapes. By contrast, our method, which takes into account both long and short range flow information, converges well at point cloud surfaces while also achieving the best CD results. We produce the best filtered results on almost all metrics. Most notably, on PUNet data with 50K points and Gaussian noise of $\sigma=3\%$, we observe a 3\% reduction in Chamfer distance error and a 6\% decrease in Point2Mesh errors, compared to StraightPCF. Moreover, on the PCNet data, our method has a clear advantage over PD-LTS.


\begin{figure*}[!ht]
    \centering
    \includegraphics[width=\linewidth]{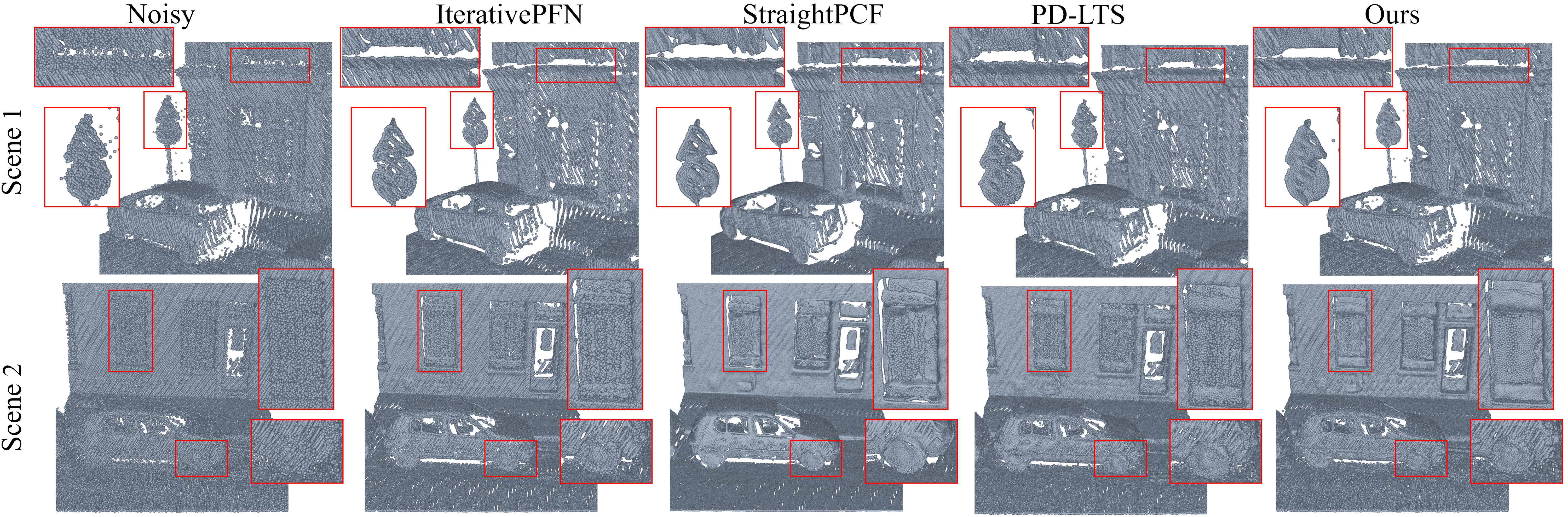}
    \caption{Visualization of filtering results on two scenes of the Paris-Rue-Madame dataset. We observe more significant clustering artifacts produced by other methods, as evident by the holes in the sign post (scene 1). Moreover, our method outperforms others in filtering surface details such as the window in scene 2.} 
    \label{fig:rue-madame-results}
\end{figure*}

\subsection{Evaluation results on real-world data}
On the Paris-Rue-Madame data, we offer more significant improvements, compared to the state-of-the-art methods. As illustrated by Fig.~\ref{fig:rue-madame-results}, our method recovers surfaces more accurately, such as the sign-post in scene 1, where other methods leave large holes. Moreover, the region connecting the wall and ledge is better filtered, without a large hole (in the region occluded from the scanner), unlike PD-LTS and StraightPCF results. Scene 2 further showcases our method's superiority. We recover a better distributed set of points on the car tire while also obtaining a smoother surface at the window without any significant holes. Next we consider quantitative filtering performance on real-world scanned data. Table~\ref{tab:kinect-results} illustrates results on Kinect data which possess a unique noise pattern. Our method performs competitively, obtaining the best result on the Chamfer distance metric. \dt{Finally, we evaluate the runtimes of different state-of-the-art methods, w.r.t. ours. We see runtime improvements over other displacement based methods, as evident in Table~\ref{tab:runtimes}.}

\midsepremove
\begin{table}[!htb]
      \centering
        \caption{Results on Kinect data.} CD and P2M values are multiplied by $10^4$.
        \vspace{1pt}
        \begin{tabular}{l|ll}
        \toprule
        \multicolumn{1}{c|}{\multirow{2}{*}{Method}} & \multicolumn{2}{c}{Kinect} \\
        \cmidrule{2-3}
         & CD & P2M \\
        \midrule
        PDFlow \cite{Mao-PDFlow} & 1.334 & 0.699 \\
        IterativePFN \cite{Edirimuni-IterativePFN} & 1.320 & 0.685 \\
        StraightPCF \cite{Edirimuni-StraightPCF} & 1.296 & \bf 0.664 \\
        PD-LTS \cite{Mao-PD-LTS} & 1.351 & 0.722 \\
        \midrule
        \bf Ours & \bf 1.285 & \uline{0.675} \\
        \bottomrule
        \end{tabular}
        \label{tab:kinect-results}
\end{table}
\midsepdefault

\midsepremove
\begin{table}[!htb]
      \centering
        \caption{\dt{We provide filtering runtimes for PUNet data at 50K resolution and 3\% Gaussian noise. We see a considerable runtime improvement w.r.t. other displacement based methods \cite{Edirimuni-IterativePFN, Edirimuni-StraightPCF}.} }
        \vspace{3pt}
        \begin{tabular}{l|ll|ll}
        \toprule
        Method & Time (s) \\
        \midrule        
        PDFlow \cite{Mao-PDFlow} & 53.8 \\
        IterativePFN \cite{Edirimuni-IterativePFN} & 19.7 \\
        StraightPCF \cite{Edirimuni-StraightPCF} & 27.04 \\
        PD-LTS \cite{Mao-PD-LTS} & 13.46 \\
        \midrule
        \bf Ours & 15.56 \\
        \bottomrule
        \end{tabular}
        \label{tab:runtimes} 
\end{table}
\midsepdefault

\section{Ablation studies}
\label{sec:ablation}
To investigate the effectiveness of our hybrid point filtering mechanism, we conduct the following ablations: we 1) compare with a baseline score model without the long range flow conditioning (BaselineScore), 2) swap the the Dynamic EdgeConv layers in the decoder for a simpler decoder with only fully connected layers (HybridPFv1), and 3) we examine the impact of using $\mathcal{L}^{EMD}_S$ vs. $\mathcal{L}^{L2}_S$ (HybridPFv2). All HybridPF variants and the BaselineScore model have roughly the same number of parameters, $\sim 480K$, similar to our final HybridPF architecture. Table~\ref{tab:ablation} shows the ablation results. 
We see a sizeable difference in results between HybridPF and the BaselineScore model, even though both networks have the same size. This validates our insight that LongModule features can positively impact short range score estimation. Moreover, as noise level increases, the separation in results also increases. Furthermore, our introduced Dynamic EdgeConv based decoder has a positive effect on filtering, with noticeable improvements to performance across all settings. Finally, we see that using $\mathcal{L}^{EMD}_S$ to supervise the ShortModule training improves overall filtering results, compared to $\mathcal{L}^{L2}_S$. We provide further ablations on the $\lambda$ hyperparameter, for loss supervision, and filtering iteration number, in the supplementary. 

\midsepremove
\begin{table}[!t]
\centering
\caption{Ablation results for showcasing the importance of 1) the LongModule, 2) introduced decoder utilizing Dynamic EdgeConv layers and 3) the choice of loss function. }
\setlength\tabcolsep{3pt} 
\begin{tabular}{l|ll|ll|ll}
\toprule
\multicolumn{1}{c|}{\multirow{2}{*}{PUNet 10K points}} & \multicolumn{2}{c|}{1\% noise} & \multicolumn{2}{c|}{2\% noise} & \multicolumn{2}{c}{3\% noise} \\
\cmidrule{2-7}
& CD & P2M & CD & P2M & CD & P2M \\ 
\midrule
BaselineScore & 1.97 & 0.30 & 2.68 & 0.61 & 3.54 & 1.35 \\
HybridPFv1 w/ FC & 1.85 & 0.22 & 2.63 & 0.56 & 3.34 & 1.15 \\
HybridPFv2 w/ $\mathcal{L}^{L2}_S$  & 1.88 & 0.25 & 2.76 & 0.69 & 3.36 & 1.19 \\
\midrule
HybridPF & \bf 1.83 & \bf 0.21 & \bf 2.55 & \bf 0.52 & \bf 3.23 & \bf 1.11 \\
\bottomrule
\end{tabular}
\label{tab:ablation} 
\end{table}
\midsepdefault

\section{Limitations and Future Work}
\label{sec:limitations}
Our method shows an impressive ability to recover good point distributions which are close to the the original clean points, even better than recent IterativePFN or PD-LTS. This is especially true on real-world Paris-Rue-Madame data, showcasing excellent generalizability to unseen data. However, we observe that on Kinect data, our method performs sub-optimally on the P2M metric despite showcasing the best CD result. We suspect this is mainly due to the Kinect noise pattern, which is quite different from synthetic Gaussian noise and LiDAR scan noise (manifested in the Paris-Rue-Madame data). In future work, we hope to extend the HybridPF's long and short range flow estimation mechanisms to account for such different noise patterns.

\section{Conclusion}
\label{sec:conclusion}
In this paper, we propose a unique hybrid method that considers both short range score and constant long range flow information when removing noise from points. To the best of our knowledge, this is the first work to consider such a hybrid short-and-long range flow mechanism for point cloud filtering. We also introduce a joint training scheme to train both the ShortModule and LongModule simultaneously, simplifying the model training process. Finally, we make improvements to decoder architecture that results in sizable advantages. Our method shows strong results on standard metrics across different datasets where we exhibit superior results to state-of-the-art methods, indicating its robustness. It also demonstrates great generalizability to unseen data such as LiDAR scan data.

\bibliographystyle{IEEEtran}
\bibliography{IEEEabrv,main}
\vskip -2\baselineskip plus -1fil
\begin{IEEEbiography}[{\includegraphics[width=1in,height=1.25in,clip,keepaspectratio]{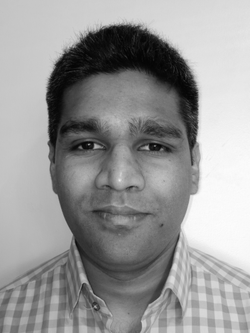}}]{Dasith de Silva Edirimuni}
received the BSc degree in Physics from Hardin-Simmons University, Texas, USA in 2013 and the MSc degree in Physics from the University of Melbourne, Australia, in 2017. In 2020 he received the Master of Information Technology degree from Swinburne University, Australia and completed his PhD in Information Technology at Deakin University, Australia in 2024. He has been a post-doctoral researcher at the University of Western Australia since August, 2024. His research interests include 3D computer vision areas such as point cloud filtering, generation and scene understanding.
\end{IEEEbiography}
\vskip -2\baselineskip plus -1fil
\begin{IEEEbiography}[{\includegraphics[width=1in,height=1.25in,clip,keepaspectratio]{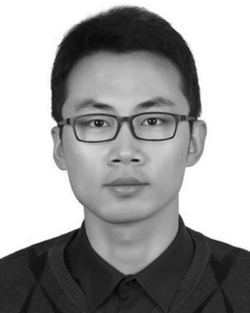}}]{Xuequan Lu}
is the Director of The Visual Computing Group (TVCG), and a Senior Lecturer at the Department of Computer Science and Software Engineering, The University of Western Australia, Australia. He spent more than two years as a Research Fellow in Singapore. Prior to that, he earned his PhD at Zhejiang University (China) in 2016. His research interests mainly fall into visual computing, e.g., 3D vision/graphics, VR/AR, 2D image processing, and digital health. More information can be found at \url{http://www.xuequanlu.com}.
\end{IEEEbiography}
\vskip -2\baselineskip plus -1fil
\begin{IEEEbiography}[{\includegraphics[width=1in,height=1.25in,clip,keepaspectratio]{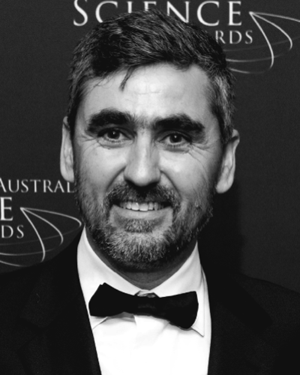}}]{Ajmal Saeed Mian}
(Senior Member, IEEE) is currently
a Professor of Computer Science with The University of Western Australia, Perth, WA, USA. He received three esteemed national fellowships and several major research grants from the Australian Research Council, the National Health and Medical Research Council, and the U.S. Department of Defense DARPA. His research interests include 3-D computer vision, machine learning, point cloud analysis, human action, and video analysis. Mr. Mian is a fellow of the International Association for Pattern Recognition. He received several awards including the
HBF Mid-Career Scientist of the Year Award in 2022 and the West Australian Early Career Scientist of the Year Award in 2012. He serves as a Senior Editor for IEEE Transactions on Neural Networks and Learning Systems and an Associate Editor for Pattern Recognition journal.
\end{IEEEbiography}
\vskip -2\baselineskip plus -1fil
\begin{IEEEbiography}[{\includegraphics[width=1in,height=1.25in,clip,keepaspectratio]{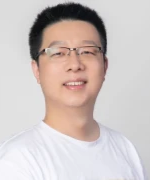}}]{Lei Wei}
(Senior Member, IEEE) received his Ph.D. in 2011 from Fraunhofer IDM@NTU in Singapore, a joint research centre between Fraunhofer-Gesellschaft and Nanyang Technological University. Between 2011 and 2019, Prof. Wei worked as a tenured Senior Researcher at the largest haptic research group in the Southern Hemisphere - Institute for Intelligent Systems Research and Innovation (IISRI), conducting research on haptics, surgical robotics and VR/AR/MR. Between 2019 and 2022, Prof. Wei worked at Tencent Robotics X Lab as a Principal Research Scientist, a Project Manager, and Team Manager of the Center for Sensing, Perception and Interaction. Prof. Wei returned to academia in mid-year 2023 as an Associate Professor at Deakin University.
\end{IEEEbiography}
\vskip -2\baselineskip plus -1fil
\begin{IEEEbiography}[{\includegraphics[width=1in,height=1.25in,clip,keepaspectratio]{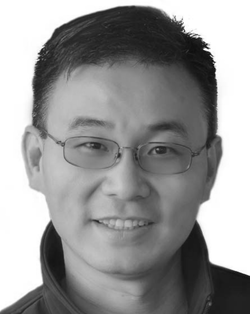}}]{Gang Li}
(Senior Member, IEEE) is currently an
Associate Professor with the School of Information
Technology, Deakin University, Melbourne, VIC,
Australia. His research interests include data mining, data privacy, causal discovery, and business
intelligence.
\end{IEEEbiography}
\vskip -2\baselineskip plus -1fil
\begin{IEEEbiography}[{\includegraphics[width=1in,height=1.25in,clip,keepaspectratio]{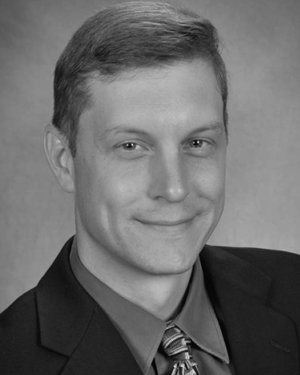}}]{Scott Schaefer} received the bachelor's degree in computer science/mathematics from Trinity University, in 2000, and the MS and PhD degrees in computer science from Rice University, in 2003 and 2006, respectively. He is currently a professor of Computer Science with Texas A\&M University. His research interests include graphics, geometry processing, curve and surface representations, and barycentric coordinates. He received the G\"{u}nter Enderle Award in 2011 and an NSF CAREER Award in 2012.
\end{IEEEbiography}
\vskip -2\baselineskip plus -1fil
\begin{IEEEbiography}[{\includegraphics[width=1in,height=1.25in,clip,keepaspectratio]{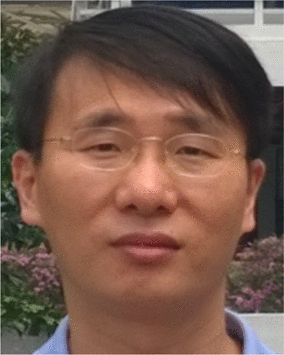}}]{Ying He}
(Senior Member, IEEE) received the B.S. and M.S. degrees in electrical engineering from Tsinghua University, Beijing, China, and the Ph.D. degree in computer science from the State University of New York, Stony Brook, NY, USA.,He is currently an Associate Professor with the School of Computer Science and Engineering, Nanyang Technological University, Singapore. His current research interest includes visual computing, with a focus on the problems that require geometric computation and analysis. .
\end{IEEEbiography}
\vskip -2\baselineskip plus -1fil

\setcounter{section}{0}
\renewcommand\thesection{\Alph{section}}
\renewcommand\thesubsection{\thesection.\arabic{subsection}}

\title{Supplementary Document for Hybrid Long and Short Range Flows for Point Cloud Filtering}

\author{Dasith~de~Silva~Edirimuni,~
        Xuequan~Lu,~\IEEEmembership{Senior~Member,~IEEE,}
        Ajmal~Saeed~Mian,~\IEEEmembership{Senior~Member,~IEEE,}
        Lei~Wei,~
        Gang~Li,~\IEEEmembership{Senior~Member,~IEEE,}
        Scott~Schaefer,~
        and~Ying~He
\IEEEcompsocitemizethanks{\IEEEcompsocthanksitem D. de Silva Edirimuni, X. Lu and A. Mian are with the Department of Computer Science and Software Engineering, University of Western Australia, Crawley, Western Australia, 6009, Australia (e-mail: \{dasith.desilva, bruce.lu, ajmal.mian\}@uwa.edu.au). L. Wei and G. Li are with the School of Information Technology, Deakin University, Waurn Ponds, Victoria, 3216, Australia (e-mail: \{lei.wei, gang.li\}@deakin.edu.au). S. Schaefer is with the Department of Computer Science and Engineering, Texas A\&M University, College Station, TX, USA (e-mail: schaefer@cse.tamu.edu). Y. He is with the College of Computing and Data Science, Nanyang Technological University, Singapore (e-mail: yhe@ntu.edu.sg).}
\thanks{Manuscript received Month Day, Year; revised Month Day, Year.}
\thanks{(Corresponding author: Xuequan Lu.)}}

\markboth{Journal of \LaTeX\ Class Files,~Vol.~14, No.~8, August~2021}%
{Shell \MakeLowercase{\textit{et al.}}: A Sample Article Using IEEEtran.cls for IEEE Journals}

\IEEEpubid{0000--0000/00\$00.00~\copyright~2021 IEEE}

\maketitle

Here we first provide additional ablation studies related to the choice of method hyperparameters and network training. Then we present additional details for filtering results on different noise patterns.

\section{Further Ablations}
In this section, we present further ablations that were not presented in the main paper.

\subsection{Discussion on the discretization parameter $\alpha$ and iteration number $N$}
We find that the discretization parameter $\alpha$ in Eq. (10) is extremely important to filtering accuracy. In general our HybridPF architecture performs well when filtering point clouds with $\sigma<2\%$ noise. However, at larger noise scales, we see a drop off in performance. Methods such as ScoreDenoise and StraightPCF~\cite{Luo-Score-Based-Denoising, Edirimuni-StraightPCF} deal with high noise by repeating the entire filtering run many times over. That is, they apply $N$ iterations of filtering $M$ times such that the point cloud is filtered a total of $T=N\times M$ times. We observe that for HybridPF, trained with the Earth Mover's Distance based short range loss function, the network overspecializes on the training noise scales, where the maximum noise level is $\sigma=2\%$. Therefore, for filtered displacements of point clouds at higher noise, we can simply vary $\alpha$ at test time to scale the inferred displacements. In general, for $0\%\leq\sigma\leq2\%$, we set $\alpha=0.8$ based on empirical results. This procedure is similar to the way in which ScoreDenoise~\cite{Luo-Score-Based-Denoising} and DeepPSR~\cite{Chen-DeepPSR} obtain $\alpha$. For $\sigma=3\%$ and 10K resolution, we set $\alpha=1.3$. Similarly, at $\sigma=3\%$ and 50K resolution, we set $\alpha=1.5$. By appropriately scaling the filtered displacements, we are not required to repeat the entire filtering run, leading to performance speedups.

\midsepremove
\begin{table}[!ht]
\centering
\caption{Ablation results for different iteration numbers $N$. }
\setlength\tabcolsep{3pt} 
\begin{tabular}{l|ll|ll|ll}
\toprule
\multicolumn{1}{c|}{\multirow{2}{*}{PUNet 10K points}} & \multicolumn{2}{c|}{1\% noise} & \multicolumn{2}{c|}{2\% noise} & \multicolumn{2}{c}{3\% noise} \\
\cmidrule{2-7}
& CD & P2M & CD & P2M & CD & P2M \\ 
\midrule
HybridPF $N=1$ & 1.93 & 0.25 & 2.91 & 0.69 & 4.14 & 1.56 \\
HybridPF $N=2$ & \uline{1.84} & \bf 0.21 & 2.60 & \uline{0.53} & 3.29 & \bf 1.11 \\
HybridPF $N=3$ & \bf 1.83 & \bf 0.21 & 2.56 & \bf 0.52 & \uline{3.24} & \bf 1.11 \\
HybridPF $N=6$ & 1.85 & \uline{0.22} & \bf 2.54 & \bf 0.52 & \uline{3.24} & \uline{1.13} \\
HybridPF $N=8$ & 1.89 & 0.24 & \uline{2.55} & 0.54 & 3.29 & 1.17 \\
\midrule
HybridPF $N=4$ & \bf 1.83 & \bf 0.21 & \uline{2.55} & \bf 0.52 & \bf 3.23 & \bf 1.11 \\
\bottomrule
\end{tabular}
\label{tab:n-ablation} 
\end{table}
\midsepdefault

Next, we consider the number of filtering iterations, $N$, in Eq. (10). Table \ref{tab:n-ablation} provides quantitative results when varying $N$. We see that a minimum of $N=2$ filtering iterations are required to obtain strong results. Moreover, the $N=4$ setting provides optimal results at both low and high noise levels, while striking a good balance between accuracy and speed.

\IEEEpubidadjcol

\subsection{Choice of loss hyperparameter}
\midsepremove
\begin{table}[!h]
\centering
\caption{Ablation results on the loss hyperparameter $\lambda$. }
\setlength\tabcolsep{3pt} 
\begin{tabular}{l|ll|ll|ll}
\toprule
\multicolumn{1}{c|}{\multirow{2}{*}{PUNet 10K points}} & \multicolumn{2}{c|}{1\% noise} & \multicolumn{2}{c|}{2\% noise} & \multicolumn{2}{c}{3\% noise} \\
\cmidrule{2-7}
& CD & P2M & CD & P2M & CD & P2M \\ 
\midrule
HybridPF $\lambda=2.0$ & \bf 1.83 & \bf 0.21 & 2.57 & 0.53 & 3.26 & \uline{1.12} \\
HybridPF $\lambda=5.0$ & 1.87 & 0.23 & 2.59 & 0.55 & 3.27 & 1.13 \\
HybridPF $\lambda=20.0$ & \uline{1.85} & \uline{0.22} & \uline{2.56} & \uline{0.53} & \uline{3.24} & \bf 1.11 \\
\midrule
HybridPF $\lambda=10.0$ & \bf 1.83 & \bf 0.21 & \bf 2.55 & \bf 0.52 & \bf 3.23 & \bf 1.11 \\
\bottomrule
\end{tabular}
\label{tab:lambda-ablation} 
\end{table}
\midsepdefault

In Eq. (9), the hyperparameter $\lambda$ weights the short range loss term w.r.t. the long range loss term. Table~\ref{tab:lambda-ablation} provides results on the effect of varying $\lambda$ on our overall pipeline. We see that $\lambda=10.0$ strikes the best balance between short range and long range loss contributions when training the HybridPF architecture.

\midsepremove
\begin{table*}[!h]
\setlength{\tabcolsep}{5.5pt}
\centering
\caption{Filtering results on the PUNet and PCNet datasets with non-isotropic Gaussian noise. This demonstrates the generalization capability to unseen noise patterns. Similar to the case of Gaussian noise, our method consistently achieves best or second-best results across nearly all settings. CD and P2M values are multiplied by $10^4$. }
\begin{tabular}{c|l| cccccc|cccccc}
\toprule
\multicolumn{2}{c|}{Resolution} & \multicolumn{6}{c|}{10K (Sparse)} & \multicolumn{6}{c}{50K (Dense)} \\
\midrule
\multicolumn{2}{c|}{Noise} & \multicolumn{2}{c}{1\%} & \multicolumn{2}{c}{2\%} & \multicolumn{2}{c|}{3\%} & \multicolumn{2}{c}{1\%} & \multicolumn{2}{c}{2\%} & \multicolumn{2}{c}{3\%} \\
\midrule
\multicolumn{2}{c|}{Method} & CD & P2M & CD & P2M & CD & P2M & CD & P2M & CD & P2M & CD & P2M \\
\midrule
\parbox[t]{2mm}{\multirow{6}{*}{\rotatebox[origin=c]{90}{PUNet~\cite{Yu-PUNet}}}}
    & Score~\cite{Luo-Score-Based-Denoising}     & 2.48 & 0.46 & 3.70 & 1.10 & 4.76 & 1.98 & 0.71 & 0.15 & 1.32 & 0.59 & 2.09 & 1.16 \\
    & PointFilter~\cite{Zhang-Pointfilter}       & 2.40 & 0.43 & 3.53 & 0.89 & 5.24 & 2.00 & 0.76 & 0.19 & 0.96 & 0.30 & 1.92 & 0.96 \\
    & PDFlow~\cite{Mao-PDFlow}                   & 2.10 & 0.38 & 3.29 & 1.06 & 4.64 & 2.18 & 0.65 & 0.17 & 1.20 & 0.60 & 2.08 & 1.35 \\
    & DeepPSR~\cite{Chen-DeepPSR}                & 2.38 & 0.33 & 3.40 & 0.81 & 4.18 & 1.38 & 0.64 & 0.08 & 1.04 & 0.35 & \bf 1.52 & \bf 0.69 \\
    & IterativePFN~\cite{Edirimuni-IterativePFN} & 1.99 & \bf 0.21 & 3.03 & 0.57 & 4.61 & 1.68 & 0.60 & \bf 0.06 & 0.83 & \bf 0.21 & 2.43 & 1.38 \\
    & StraightPCF~\cite{Edirimuni-StraightPCF}   & 1.83 & 0.24 & 2.64 & 0.62 & \uline{3.48} & \uline{1.29} & 0.56 & \uline{0.11} & 0.80 & 0.30 & 1.66 & 0.93 \\
    & PD-LTS~\cite{Mao-PD-LTS}                   & \bf 1.80 & \bf 0.21 & \uline{2.59} & \bf 0.55 & 3.62 & \uline{1.29} & \uline{0.50} & \bf 0.06 & \bf 0.74 & \uline{0.22} & 1.85 & 1.06 \\
\cmidrule{2-14}
    & Ours                                       & \uline{1.81} & \bf 0.21 & \bf 2.58 & \uline{0.56} & \bf 3.43 & \bf 1.27 & \bf 0.49 & \bf 0.06 & \uline{0.76} & 0.25 & \uline{1.56} & \uline{0.84} \\
\midrule
\parbox[t]{2mm}{\multirow{6}{*}{\rotatebox[origin=c]{90}{PCNet~\cite{Rakotosaona-PCN}}}}
    & Score~\cite{Luo-Score-Based-Denoising}     & 3.38 & 0.83 & 5.17 & 1.22 & 6.82 & 1.98 & 1.08 & 0.18 & 1.74 & 0.39 & 2.69 & 0.76 \\
    & PointFilter~\cite{Zhang-Pointfilter}       & 3.02 & 0.88 & 4.95 & 1.33 & 7.48 & 2.23 & 1.07 & 0.19 & 1.46 & 0.31 & 2.72 & 0.69 \\
    & PDFlow~\cite{Mao-PDFlow}                   & 3.25 & 0.61 & 4.72 & 0.99 & 6.39 & 1.74 & 0.99 & 0.16 & 1.73 & 0.48 & 2.83 & 0.81 \\
    & DeepPSR~\cite{Chen-DeepPSR}                & 3.15 & 0.99 & 4.94 & 1.34 & 6.31 & 1.86 & 1.01 & 0.17 & 1.63 & 0.43 & 2.33 & 0.72 \\
    & IterativePFN~\cite{Edirimuni-IterativePFN} & \uline{2.64} & 0.69 & 4.49 & 1.04 & 6.53 & 1.91 & 0.92 & \uline{0.14} & 1.32 & \uline{0.27} & 3.05 & 0.99 \\
    & StraightPCF~\cite{Edirimuni-StraightPCF}   & \uline{2.76} & \uline{0.54} & \uline{4.08} & \uline{0.83} & \bf 5.21 & \bf 1.33 & 0.88 & 0.15 & \bf 1.23 & 0.29 & \bf 2.23 & \bf 0.65 \\
    & PD-LTS~\cite{Mao-PD-LTS}                   & 2.88 & 0.55 & 4.26 & 0.91 & 5.65 & 1.49 & \uline{0.87} & \uline{0.14} & 1.34 & 0.31 & 2.55 & 0.79 \\
\cmidrule{2-14}
    & Ours                                       & 2.83 & \bf 0.48 & \bf 4.07 & \bf 0.76 & \uline{5.39} & \uline{1.35} & \bf 0.80 & \bf 0.10 & \uline{1.25} & \bf 0.25 & \uline{2.26} & \uline{0.67} \\
\bottomrule
\end{tabular}
\label{tab:pu-pc-cov-results}
\end{table*}
\midsepdefault

\begin{figure*}[!h]
\centering
\includegraphics[width=0.97\linewidth]{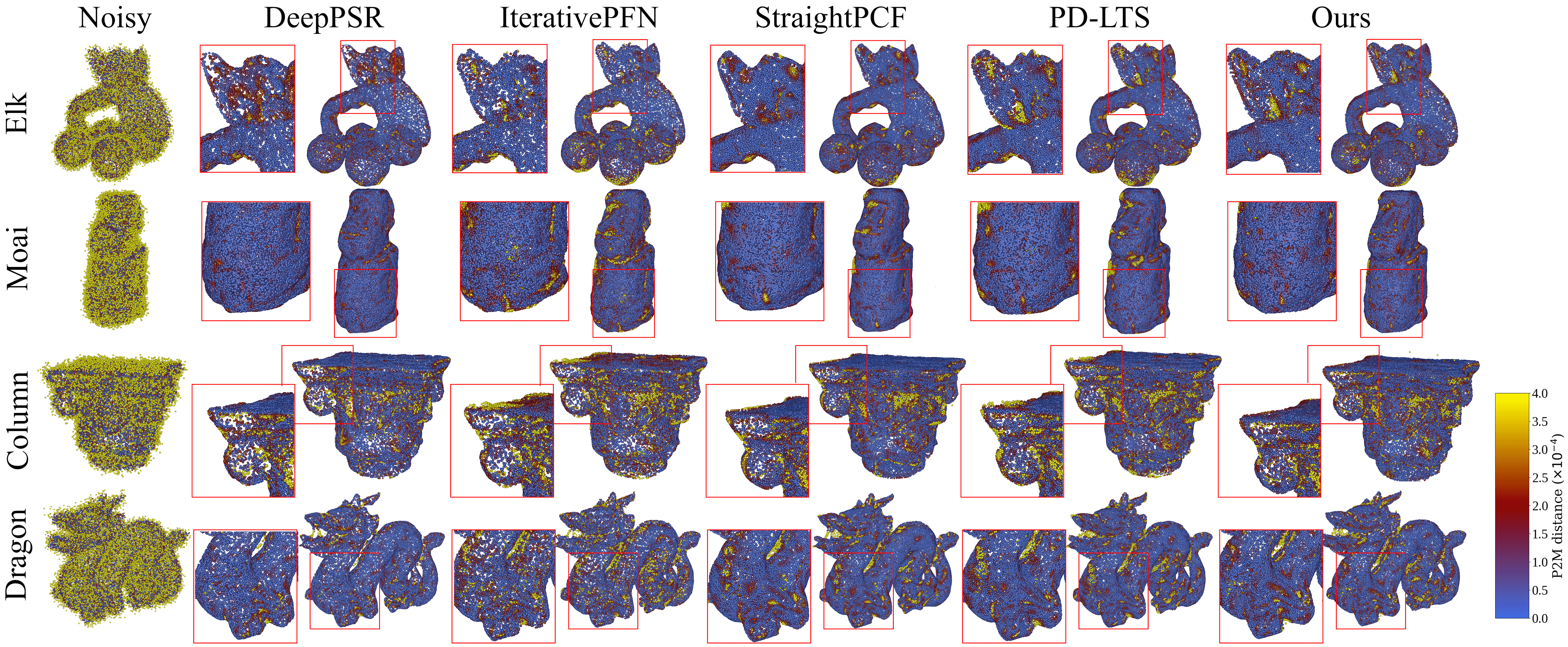}
\caption{Heatmap visualization of filtered point clouds on PUNet and PCNet data at high resolution and high non-isotropic Gaussian noise (i.e., 50K resolution shapes with $\sigma=3\%$). Blue indicates filtered points are closer to the clean surface. }
\label{fig:pu-pc-cov-results}
\end{figure*}

\section{Filtering Results on Different Noise Patterns}
\subsection{Non-isotropic Gaussian noise}
The main paper results focused on synthetic data where the Gaussian noise pattern is isotropic. Therefore, we also investigate the case of non-isotropic Gaussian noise, to demonstrate generalization of our method to noise patterns beyond that of the initial training data. The covariance matrix $\Sigma$ for this non-isotropic noise pattern is given by:
\begin{equation}\label{eq:non-isotropicnoise}
    \Sigma = \sigma^2\times \begin{bmatrix}
                        1 & -1/2 & -1/4 \\
                        -1/2 & 1 & -1/4 \\
                        -1/4 & -1/4 & 1
                        \end{bmatrix},
\end{equation}
here the noise scale $\sigma$ is set to $1\%$, $2\%$ and $3\%$ of the bounding sphere's radius. In general, as illustrated in Table~\ref{tab:pu-pc-cov-results}, our method performs more consistently than others when filtering PUNet data and shows a strong ability to recover the underlying clean point distribution correctly (given by top or second best CD results across most metrics). Furthermore, on the PCNet data, our method performs better that all methods on nearly all resolution and noise settings for both metrics. Fig.~\ref{fig:pu-pc-cov-results} demonstrate visual results where we recover filtered points while avoiding clustering, unlike DeepPSR or IterativePFN.

\subsection{Uniformly distributed noise}
\begin{figure*}[!tp]
\centering
\includegraphics[width=0.97\linewidth]{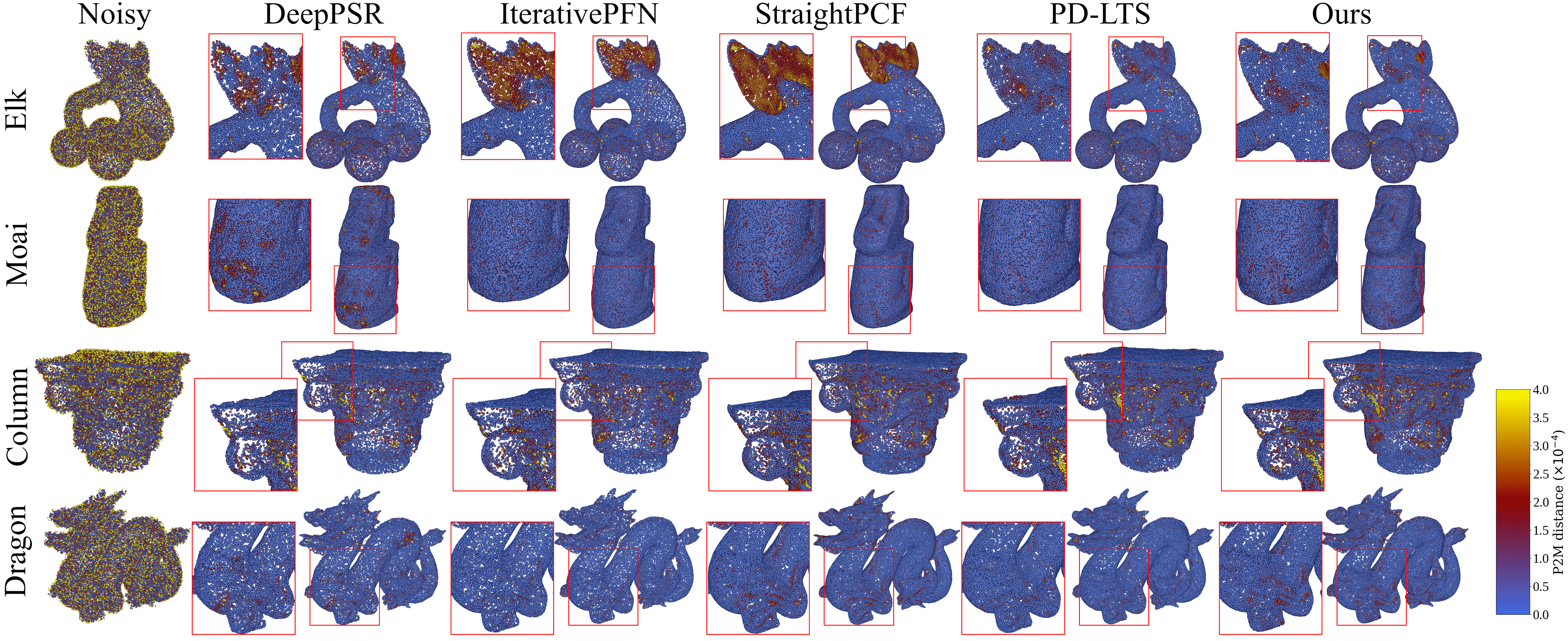}
\caption{Heatmap visualization of filtered point clouds on PUNet and PCNet data at high resolution and high uniformly-distributed noise (i.e., 50K resolution shapes with $\sigma=3\%$). Blue indicates filtered points are closer to the clean surface. }
\label{fig:pu-pc-uniform-results}
\end{figure*}

Next we consider the case of uniformly distributed noise within a sphere of radius $\sigma$. Thus, the probability to sample a point at a position $\pmb{x}$ is given by,
\begin{equation}
    p(\pmb{x}; \sigma) = \begin{cases} \frac{3}{4\pi \sigma^3}, &\norm{\pmb{x}}_2\leq \sigma, \\
    0, &\text{Otherwise}
    \end{cases}
\end{equation}
We set the sphere radius $\sigma$ to be 1\%, 2\% and 3\% of the point cloud's bounding sphere radius. The quantitative filtering results are given in Table~\ref{tab:pu-pc-uniform-results} and the visual results are illustrated in Fig.~\ref{fig:pu-pc-uniform-results}. While, on the PUNet data, we obtain the second best results across most settings, on the PCNet data, we obtain best results. Moreover, visual results on shapes such as Elk, Column and Dragon indicate that, while other methods' filtering results suffer from clustering, our filtered points have nicer distributions. 

\midsepremove
\begin{table*}[!tp]
\setlength{\tabcolsep}{5.5pt}
\centering
\caption{Filtering results on the PUNet and PCNet datasets with uniformly distributed noise. Our method performs especially well on the PCNet data. Moreover, we obtain best or second-best results across most noise and resolution settings. CD and P2M values are multiplied by $10^4$.}
\begin{tabular}{c|l| cccccc|cccccc}
\toprule
\multicolumn{2}{c|}{Resolution} & \multicolumn{6}{c|}{10K (Sparse)} & \multicolumn{6}{c}{50K (Dense)} \\
\midrule
\multicolumn{2}{c|}{Noise} & \multicolumn{2}{c}{1\%} & \multicolumn{2}{c}{2\%} & \multicolumn{2}{c|}{3\%} & \multicolumn{2}{c}{1\%} & \multicolumn{2}{c}{2\%} & \multicolumn{2}{c}{3\%} \\
\midrule
\multicolumn{2}{c|}{Method} & CD & P2M & CD & P2M & CD & P2M & CD & P2M & CD & P2M & CD & P2M \\
\midrule
\parbox[t]{2mm}{\multirow{6}{*}{\rotatebox[origin=c]{90}{PUNet~\cite{Yu-PUNet}}}}
    & Score~\cite{Luo-Score-Based-Denoising}     & 1.27 & 0.25 & 2.47 & 0.41 & 3.37 & 0.98 & 0.51 & 0.05 & 0.69 & 0.13 & 0.90 & 0.29 \\
    & PointFilter~\cite{Zhang-Pointfilter}       & 1.14 & 0.29 & 2.45 & 0.41 & 3.03 & 0.57 & 0.63 & 0.15 & 0.74 & 0.17 & 0.80 & 0.19 \\
    & PDFlow~\cite{Mao-PDFlow}                   & 0.87 & 0.18 & 2.03 & 0.33 & 2.66 & 0.64 & 0.46 & 0.06 & 0.85 & 0.33 & 0.95 & 0.40 \\
    & DeepPSR~\cite{Chen-DeepPSR}                & 1.23 & 0.19 & 2.42 & 0.30 & 2.97 & 0.47 & 0.50 & 0.03 & 0.64 & 0.06 & 0.93 & 0.25 \\
    & IterativePFN~\cite{Edirimuni-IterativePFN} & 0.65 & \uline{0.09} & 2.01 & 0.19 & 2.68 & \uline{0.34} & 0.44 & \bf 0.01 & 0.60 & \bf 0.05 & 0.69 & \bf 0.11 \\
    & StraightPCF~\cite{Edirimuni-StraightPCF}   & \uline{0.62} & 0.10 & 1.83 & 0.25 & 2.37 & 0.43 & 0.42 & 0.05 & \uline{0.56} & 0.13 & 0.68 & 0.23 \\
    & PD-LTS~\cite{Mao-PD-LTS}                   & \bf 0.61 & 0.08 & 1.76 & \bf 0.18 & \bf 2.26 & \bf 0.31 & \bf 0.37 & \uline{0.02} & \bf 0.48 & \bf 0.05 & \bf 0.55 & \bf 0.11 \\
\cmidrule{2-14}
    & Ours                                       & 0.63 & \uline{0.09} & \uline{1.77} & \uline{0.19} & \uline{2.29} & 0.37 & \uline{0.38} & \uline{0.02} & \bf 0.48 & \uline{0.06} & \uline{0.63} & \uline{0.18} \\
\midrule
\parbox[t]{2mm}{\multirow{6}{*}{\rotatebox[origin=c]{90}{PCNet~\cite{Rakotosaona-PCN}}}}
    & Score~\cite{Luo-Score-Based-Denoising}     & 1.79 & 0.69 & 3.23 & 0.80 & 4.77 & 1.26 & 0.71 & 0.10 & 1.04 & 0.16 & 1.34 & 0.31 \\
    & PointFilter~\cite{Zhang-Pointfilter}       & 1.41 & 0.79 & 2.93 & 0.85 & 4.03 & 1.00 & 0.77 & 0.16 & 1.06 & 0.18 & 1.18 & 0.21 \\
    & PDFlow~\cite{Mao-PDFlow}                   & 2.04 & 0.51 & 3.11 & 0.58 & 3.95 & 0.79 & 0.73 & 0.10 & 1.25 & 0.33 & 1.37 & 0.36 \\
    & DeepPSR~\cite{Chen-DeepPSR}                & 1.61 & 0.95 & 3.07 & 1.01 & 4.15 & 1.13 & 0.66 & 0.11 & 0.99 & 0.16 & 1.31 & 0.27 \\
    & IterativePFN~\cite{Edirimuni-IterativePFN} & \bf 0.99 & 0.60 & \bf 2.50 & 0.67 & 3.71 & 0.81 & 0.58 & 0.09 & 0.92 & 0.13 & 1.07 & 0.19 \\
    & StraightPCF~\cite{Edirimuni-StraightPCF}   & \uline{1.40} & 0.48 & 2.92 & 0.58 & 3.68 & 0.70 & 0.68 & 0.10 & 0.68 & 0.10 & 0.91 & 0.18 \\
    & PD-LTS~\cite{Mao-PD-LTS}                   & 1.48 & \uline{0.44} & 2.72 & \uline{0.51} & \bf 3.55 & \uline{0.63} & \uline{0.60} & \uline{0.07} & \uline{0.60} & \uline{0.07} & \uline{0.82} & \uline{0.12} \\
\cmidrule{2-14}
    & Ours                                       & 1.52 & \bf 0.40 & \uline{2.68} & \bf 0.45 & \uline{3.64} & \bf 0.62 & \bf 0.59 & \bf 0.06 & \bf 0.59 & \bf 0.06 & \bf 0.77 & \bf 0.09 \\
\bottomrule
\end{tabular}
\label{tab:pu-pc-uniform-results}
\end{table*}
\midsepdefault

\midsepremove
\begin{table*}[!t]
\setlength{\tabcolsep}{5.5pt}
\centering
\caption{Filtering results on the PUNet and PCNet datasets with Laplace noise. This demonstrates the generalization capability to unseen noise patterns. On PUNet and PCNet data, our method outperforms others at 10K across most noise settings. We note that PCNet data is unseen during training. CD and P2M values are multiplied by $10^4$. }
\begin{tabular}{c|l| cccccc|cccccc}
\toprule
\multicolumn{2}{c|}{Resolution} & \multicolumn{6}{c|}{10K (Sparse)} & \multicolumn{6}{c}{50K (Dense)} \\
\midrule
\multicolumn{2}{c|}{Noise} & \multicolumn{2}{c}{1\%} & \multicolumn{2}{c}{2\%} & \multicolumn{2}{c|}{3\%} & \multicolumn{2}{c}{1\%} & \multicolumn{2}{c}{2\%} & \multicolumn{2}{c}{3\%} \\
\midrule
\multicolumn{2}{c|}{Method} & CD & P2M & CD & P2M & CD & P2M & CD & P2M & CD & P2M & CD & P2M \\
\midrule
\parbox[t]{2mm}{\multirow{6}{*}{\rotatebox[origin=c]{90}{PUNet~\cite{Yu-PUNet}}}}
    & Score~\cite{Luo-Score-Based-Denoising}     & 2.90 & 0.67 & 4.60 & 1.81 & 6.30 & 3.23 & 0.83 & 0.23 & 1.68 & 0.89 & 2.67 & 1.65 \\
    & PointFilter~\cite{Zhang-Pointfilter}       & 2.77 & 0.57 & 4.25 & 1.37 & 7.67 & 3.91 & 0.83 & 0.23 & 1.24 & 0.49 & 2.88 & 1.80 \\
    & PDFlow~\cite{Mao-PDFlow}                   & 2.54 & 0.57 & 4.32 & 1.86 & 7.82 & 4.82 & 0.82 & 0.30 & 1.51 & 0.86 & 4.53 & 3.52 \\
    & DeepPSR~\cite{Chen-DeepPSR}                & 2.73 & 0.49 & 3.92 & 1.22 & 4.98 & \uline{1.98} & 0.74 & 0.14 & 1.28 & 0.53 & \bf 1.84 & \bf 0.89 \\
    & IterativePFN~\cite{Edirimuni-IterativePFN} & 2.39 & \uline{0.32} & 3.40 & \bf 0.81 & 6.49 & 3.13 & 0.65 & \bf 0.09 & 1.00 & \bf 0.32 & 3.67 & 2.41 \\
    & StraightPCF~\cite{Edirimuni-StraightPCF}   & 2.16 & 0.34 & \bf 2.98 & 0.87 & \uline{4.57} & 2.16 & 0.60 & 0.13 & 0.97 & 0.41 & 2.54 & 1.68 \\
    & PD-LTS~\cite{Mao-PD-LTS}                   & \uline{2.13} & \uline{0.32} & 3.01 & 0.85 & 5.02 & 2.36 & \uline{0.56} & \uline{0.10} & \uline{0.99} & 0.39 & 2.61 & 1.67 \\
\cmidrule{2-14}
    & Ours                                       & \bf 2.12 & \uline{0.31} & \bf 2.96 & \uline{0.84} & \bf 4.14 & \bf 1.80 & \bf 0.55 & \uline{0.10} & \bf 0.94 & \uline{0.37} & \uline{2.00} & \uline{1.11} \\ 
\midrule
\parbox[t]{2mm}{\multirow{6}{*}{\rotatebox[origin=c]{90}{PCNet~\cite{Rakotosaona-PCN}}}}
    & Score~\cite{Luo-Score-Based-Denoising}     & 3.97 & 0.95 & 6.19 & 1.65 & 8.40 & 2.62 & 1.20 & 0.23 & 2.06 & 0.51 & 3.38 & 1.01 \\
    & PointFilter~\cite{Zhang-Pointfilter}       & 3.54 & 0.99 & 5.73 & 1.59 & 9.82 & 3.27 & 1.16 & 0.23 & 1.72 & 0.38 & 3.54 & 1.10 \\
    & PDFlow~\cite{Mao-PDFlow}                   & 3.76 & 0.71 & 5.78 & 1.40 & 9.25 & 2.87 & 1.17 & 0.21 & 2.11 & 0.54 & 4.83 & 1.45 \\
    & DeepPSR~\cite{Chen-DeepPSR}                & 3.69 & 1.13 & 5.45 & 1.57 & 7.14 & 2.17 & 1.12 & 0.23 & 1.83 & 0.48 & \uline{2.57} & \uline{0.74} \\
    & IterativePFN~\cite{Edirimuni-IterativePFN} & \uline{3.19} & 0.81 & 4.89 & 1.17 & 8.17 & 2.52 & 0.99 & 0.17 & 1.48 & \bf 0.29 & 3.80 & 1.14 \\
    & StraightPCF~\cite{Edirimuni-StraightPCF}   & \bf 3.16 & \uline{0.60} & \bf 4.47 & \uline{0.94} & \bf 6.10 & \bf 1.55 & \uline{0.94} & \uline{0.16} & \bf 1.40 & 0.33 & 2.92 & 0.89 \\
    & PD-LTS~\cite{Mao-PD-LTS}                   & 3.36 & 0.64 & 4.76 & 1.08 & 7.12 & 1.96 & 0.95 & 0.17 & 1.53 & 0.36 & 3.14 & 0.98 \\
\cmidrule{2-14}
    & Ours                                       & 3.25 & \bf 0.54 & \uline{4.55} & \bf 0.93 & \uline{6.13} & \uline{1.56} & \bf 0.88 & \bf 0.13 & \uline{1.41} & \uline{0.30} & \bf 2.47 & \bf 0.64 \\
\bottomrule
\end{tabular}
\label{tab:pu-pc-laplace-results}
\end{table*}
\midsepdefault

\begin{figure*}[!t]
\centering
\includegraphics[width=0.97\linewidth]{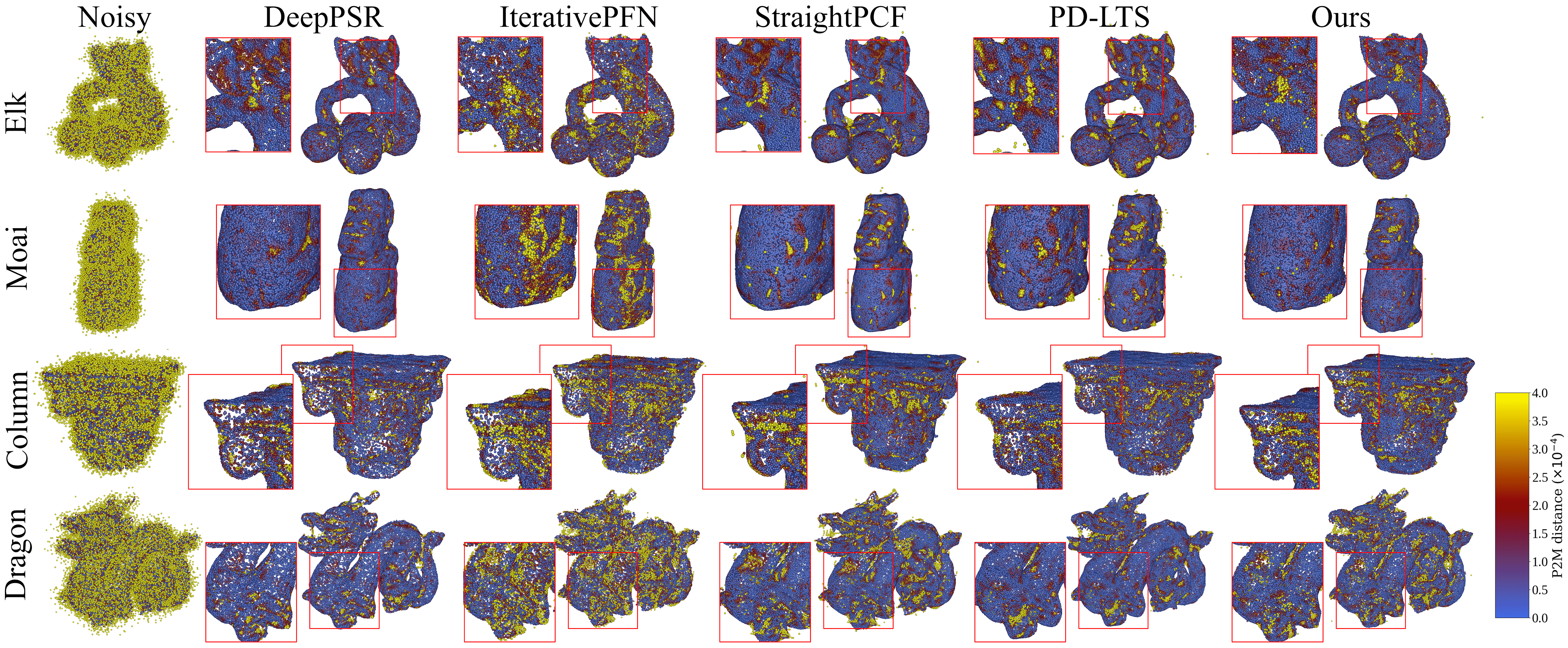}
\caption{Heatmap visualization of filtered point clouds on PUNet and PCNet data at high resolution and high Laplace noise (i.e., 50K resolution shapes with $\sigma=3\%$). Blue indicates filtered points are closer to the clean surface. }
\label{fig:pu-pc-laplace-results}
\end{figure*}

\subsection{Laplace noise}
Finally, we consider the case of Laplace noise. Similar to the evaluation settings of the main paper, we set the noise scale $\sigma$ to be 1\%, 2\% and 3\% of the point cloud's bounding sphere radius. The quantitative results are given in Table~\ref{tab:pu-pc-laplace-results} and the visual results are illustrated in Fig.~\ref{fig:pu-pc-laplace-results}. A characteristic of Laplace noise is that is at a higher noise intensity than other noise patterns. Our method performs extremely well on PUNet and PCNet data at 10K resolution. It more consistently outperforms other methods on this noise pattern, as evidenced by Table~\ref{tab:pu-pc-laplace-results}. Moreover, at high unseen noise scales of $\sigma=3\%$, HybridPF outperforms other methods across both resolution settings. This indicates the high generalizability of our method. Visual results indicate that the $\sigma=3$ setting, at 50K resolution, is challenging for all methods. However, our method better filters complex surfaces such as the Elk ears, compared to PD-LTS or StraightPCF. We also show better filtered results on shapes such as Moai, with less yellow-ish regions indicating filtered points are far from the clean surface, and Column, where our results have fewer holes, i.e., less clustering. 


\end{document}